\documentclass[journal]{IEEEtran}
\ifCLASSINFOpdf
\else
\fi
\usepackage{tikz}
\usepackage{amsmath}
\usepackage{amsfonts}

\usepackage{algorithm}  
\usepackage{algorithmic}
\usepackage{subcaption}
\usepackage{tabularx}
\usepackage{booktabs}
 
\usepackage{graphicx}
\usepackage[top=0.7in, bottom=0.7in, left=0.6in, right=0.6in]{geometry} 
\usepackage{indentfirst} 
 \usepackage{enumitem}
\begin{document}
%
\title{SPFL: A Self-purified Federated Learning Method Against Poisoning Attacks}
%
%
%

\author{Zizhen~Liu,
        Weiyang~He,
        Chip-Hong~Chang,~\IEEEmembership{Fellow,~IEEE,}
        Jing~Ye,
        Huawei~Li,~\IEEEmembership{ Senior Member,~IEEE,}
        Xiaowei~Li,~\IEEEmembership{Senior Member,~IEEE}
\vspace{-0.3cm}
\thanks{Manuscript received December 31, 2022. This research is supported in part by the Ministry of Education, Singapore, under its AcRF Tier 2 Award No MOE-T2EP20121-0001, and in part by the National Key Research and Development Program of China under grant No.2020YFB1600201, the National Natural Science Foundation of China (NSFC) under grant No. (U20A20202, 62090024, 61876173), and the Youth Innovation Promotion Association CAS.}        
\thanks{W. He and C. H. Chang are with the School of Electrical and Electronic Engineering, Nanyang Technological University, Singapore 639798. Z. Liu, Y. Jing, H. Li and X. Li are with the State Key Lab of Processors, Institute of Computing Technology, Chinese Academy of Sciences, and the University of Chinese Academy of Sciences, and CASTEST Co., Ltd., Beijing, China 100190. Corresponding author: C. H. Chang. (Emails: echchang@ntu.edu.sg)}}

%
%

\markboth{Journal of \LaTeX\ Class Files,~Vol.~14, No.~8, August~2015}%
{Shell \MakeLowercase{\textit{et al.}}: Bare Demo of IEEEtran.cls for IEEE Journals}
%



\maketitle

\begin{abstract}
While Federated learning (FL) is attractive for pulling privacy-preserving distributed training data, the credibility of participating clients and non-inspectable data pose new security threats, of which poisoning attacks are particularly rampant and hard to defend without compromising privacy, performance or other desirable properties of FL. To tackle this problem, we propose a self-purified FL (SPFL) method that enables benign clients to exploit trusted historical features of locally purified model to supervise the training of aggregated model in each iteration. The purification is performed by an attention-guided self-knowledge distillation where the teacher and student models are optimized locally for task loss, distillation loss and attention-based loss simultaneously. SPFL imposes no restriction on the communication protocol and aggregator at the server. It can work in tandem with any existing secure aggregation algorithms and protocols for augmented security and privacy guarantee. We experimentally demonstrate that SPFL outperforms state-of-the-art FL defenses against various poisoning attacks. The attack success rate of SPFL trained model is at most 3$\%$ above that of a clean model, even if the poisoning attack is launched in every iteration with all but one malicious clients in the system. Meantime, it improves the model quality on normal inputs compared to FedAvg, either under attack or in the absence of an attack.  
\end{abstract}
\begin{IEEEkeywords}
Federated Learning, Poisoning attack, Knowledge distillation, Attention maps, Deep Neural Network
\end{IEEEkeywords}

%
\IEEEpeerreviewmaketitle
\section{Introduction}
%
%
%
%
\IEEEPARstart{F}{ederated Learning}~(FL) is an emerging distributed learning paradigm for multiple participants (also known as clients) to collaboratively train a model using their own data. The workflow is typically orchestrated by a central server which collects only updates to locally trained model but not the raw training data that is stored locally and never exchanged or transferred. These updates are aggregated by the server to improve the global model performance and the updated model data is passed to the participating clients in exchange for new updates in the next iteration of training~\cite{yang2019federated}. The aggregation is usually computed in a secure manner such that each client, including the server, can learn nothing more than the aggregation result to avoid indirect privacy leakage of sensitive information for inference attacks~\cite{41zhu2019deep, 20dp}.

Beyond attacks targeting data inference, FL also opens the door to a fleet of malicious alterations of the training pipeline made feasible by distributed learning over uninspectable private datasets and privacy-preserving multi-party computation involving potentially unreliable devices. New attack surfaces may be introduced into FL system by using training pipeline as an intermediary for inference time attacks. Since individual benign clients and the server have no control over other clients' behaviors, a malicious client may deviate from the prescribed training protocol to modify the model in an undesirable way to achieve different objectives. The attacks can be conducted via inflated model updates by either crafting poisoned data for local model training (data poisoning) or directly manipulating model updates to the server (model poisoning)~\cite{1}. The attack scope can be untargeted or targeted. Untargeted attacks downgrade the model performance globally, whereas targeted poisoning attacks affect the model’s output on only a small subset of data points with specified characteristics while preserving its overall accuracy. This type of integrity threat has gained more attention recently due to the fact that it is harder to detect, and can be easily executed – a single-shot attack may suffice to create a backdoor into the centrally learned model. Some nuanced attacks may exhibit a continuum between these two outcomes. Byzantine threat model assumes that one or more adversaries can directly control certain number of clients to alter their outputs and bias the learned model to produce arbitrary outputs towards the adversarial objective. Unlike centralized learning, adversaries in FL setting have access to a pool of Byzantine devices and can inspect the model parameters to coordinate their poisoned updates in real time (within-update collusion) or adapt their attack across updates (cross-update collusion) as the training progresses. 

It is challenging to guarantee the security of FL against poisoning attacks. For one thing, the training data cannot be inspected by anyone due to data minimization. This makes existing defenses against backdoor attacks that require full control of training process, or access to training data or holdout set inapplicable in the FL setting. For another thing, if defense strategies have to access or compute on plain local gradients in defending against poisoning attacks, then local gradients are inevitably leaked, which is not permitted for privacy concern. Furthermore, adversaries are allowed to collude and control more clients at different time, which dramatically increases the effectiveness and stealth of poisoning attacks. Cross-update collusion makes both targeted and untargeted attacks more difficult to be detected and defended against as each Byzantine client is free to participate in or exit from any round of update. Moreover, defenses against model update and data poisoning attacks must assume white box adversaries in FL setting, which adds challenges to preserve privacy, robustness and model’s performance simultaneously. 

Existing defense methods based on robust aggregation rules try to replace averaging with robust estimators such as geometric median \cite{blanchard2017machine}, coordinate-wise median \cite{yin2018byzantine}, $\alpha$-trimmed mean, or a variant/combination of such techniques \cite{pillutla2022robust, 18fung2018mitigating}. Although some of these robust aggregation algorithms are provably effective against data and untargeted model poisoning attacks under certain assumptions, their theoretical guarantees are often not met on more rigorous empirical evaluation with different learning problems.  most of these Byzantine-tolerant aggregation schemes and other server-side mitigation approaches generally assume that the server is honest. They operate on visible model updates, which are not compatible or cannot be easily integrated with multi-party computation protocols that possess strong user privacy guarantee. A recent defense based on black-box knowledge transfer~\cite{chang2019cronus} reduces information leakage by reducing the dimension of the exchanged information. It shares predictions instead of model parameters of individual client models on public data.  The assumption of a large common pool of training data itself violates the central telnet of distributed learning. Using public training data to share confidence scores also exposes the local model to membership inference attack. Without a centralized model, it becomes harder to pre-empt the hyperparameters for differential privacy protection without sacrificing the accuracy of benign client models.  

It is daunting if not impractically costly to design a foolproof FL system with ideal privacy and security properties. To prevent perfect from being the enemy of good, it is more important to address security and privacy problems of FL by a divide-and-conquer approach with modular protection methods. Each mitigation method can coexist with other countermeasures to mitigate different threats and can be improved independently without hurting the security and privacy guarantees of others to minimize trade off requirement on multiple desirable properties. With this in mind, we propose a novel self-purified FL (SPFL) method by using self-distillation to defend against poisoning attacks with high Byzantine tolerance. SPFL can be performed by each benign client independently and co-exist with any secure and privacy-preserving aggregation methods. The local model of last iteration plays the teacher role, and purifies the aggregated malicious model via knowledge distillation in each iteration of FL training. An effective mechanism is proposed to transfer class-discriminating attention information between the teacher and the student. Specifically, our major contributions are:
\begin{enumerate}
[leftmargin=*] 
\item We introduce the first attention guided self-distillation technique into FL to defend against poisoning attacks, which enables benign clients to fully exploit local training data and historical model to purify malicious updates regardless of the number of colluding adversaries.  
\item The anomaly is mitigated by optimizing the overall task loss, distillation loss and attention loss functions. Class-discriminating attention information in addition to traditional logit information are exploited for teacher-student collaborative knowledge distillation,  making SPFL a powerful active client-side defense method against poisoning attack for FL.
\item We evaluate SPFL on a suite of federated benchmarks across multiple target poisoning attacks, and demonstrate that it is robust against poisoning attacks while improving the model quality concurrently. 
\item SPFL makes no assumption about trusted server and public data. It allows the benign clients to implement the defense without any restriction on or conflict with existing aggregation algorithms and secure aggregation protocols for augmented security and privacy risk mitigation.
\end{enumerate}
\section{Background and Related Work}
\subsection{Federated Learning (FL)}
FL is a distributed machine learning paradigm for building a model based on training data of multiple decentralized participants~\cite{yang2019federated}. Instead of pooling raw training data, the global model is established by collecting focused model updates from clients who trained the shared model locally using their own data for privacy preservation. Focused updates are narrowly scoped to contain minimum information for specific learning task, and they can be instantiated as model parameters, gradients or the gaps before and after model training in wide-ranging algorithms. In horizontal FL, different clients have different data but the local training datasets share the same feature and label space. Usually, the training process is orchestrated by a central server. One of the typical optimization algorithms for updating the global model is Federated Averaging (FedAvg) \cite{40kairouz2021advances}. For each communication round (iteration), each client applies stochastic gradient descent (SGD) to train the model with its local data for a number of internal epochs. The clients then transmit their local model parameters to the central server, where the global model is updated with the aggregation of the clients’ local model updates. This process is repeated until the loss function converges. As the shared model parameters carry extensive information about the corresponding training data of the clients~\cite{41zhu2019deep}, secure aggregation protocols are integrated into FedAvg to prevent indirect leakage of personal information. Secure aggregation enables multiple clients to jointly aggregate the model updates but no party can learn anything beyond the aggregation results \cite{43bonawitz2017practical, aono2017privacy}. To avoid shared model parameters from being stealthily extracted and exploited for inference attacks, individual local model updates are protected by privacy-preserving and cryptographic techniques such as homomorphic encryption~\cite{liu2022dhsa, aono2017privacy} or mask hiding~\cite{liu2022sash,43bonawitz2017practical} before they are uploaded to the server for aggregation.
\subsection{Poisoning Threats of Federated Learning}
Poisoning attack is one of the most notorious threats of FL. It corrupts a target model at the training process to impede the model from converging to the optimal performance \cite{shejwalkar2021manipulating} or control its prediction results \cite{1}. The distributed training attributes and privacy-preserving aggregation of FL infrastructure bolster such attacks. Since the model is trained locally before the updated local model parameters are submitted and aggregated by the server, any participant who has access to the local model can deviate from the prescribed protocol to conduct a training-time attack \cite{yang2019federated}. Furthermore, due to the confidentiality requirements, all participants are not supposed to know how other clients generate their updates. The coordination server is constrained by the opacity of the updates and training process at the local premises to verify the received data integrity. This gives a group of malicious participants the opportunity to directly and cooperatively manipulate the training processes or trained parameters of their local models to stealthily insert a backdoor into the global model. This way of feeding the aggregator with rigged updates from one or more local backdoored models is known as model poisoning \cite{1, 5}. It is more powerful than data poisoning attacks in centralized learning. Depending on the adversary's capability and goal, model poisoning can also be combined with data poisoning and executed simultaneously or individually in FL.\\ ${\rm{     }}$ ${\rm{     }}$ ${\rm{     }}$
According to the goals, poisoning attacks can be broadly divided into untargeted and targeted. Untargeted attacks compromise the trained model to cause misclassification of clean inputs. They tend to incur non-trivial degradation in learning performance or obstruct learning convergence. The attack in \cite{fang2020local} assumed that the local model parameters on compromised client devices can be manipulated during the learning process to produce a large testing error rate on the global model. However, benign participants can observe the apparent adverse global effect during the training process and take actions to mitigate this type of attack. In contrast, targeted attacks embed malicious functionality into the trained model by controlling the model’s prediction accuracy only under specific circumstances without disrupting its normal behavior. This type of attack maintains the model availability by achieving high accuracy on both its main task and the attacker chosen subtask. Backdoor attack is the most widely encountered targeted attack in FL, wherein a backdoor is surreptitiously infiltrated into the joint model to change the model’s outputs only on inputs with certain characteristics selected by the attacker. \\ ${\rm{     }}$ ${\rm{     }}$ ${\rm{     }}$
Bagdasaryan et al. \cite{1} developed a model poisoning attack (MPA) that allows any participants in FL to replace the joint model with a backdoored model. It was found in \cite{2} that the backdoored model can be trained by limiting the update norms to avoid norm clipping and evade anomaly detection. Byzantine participants can share the updates evenly and maliciously train the model with multiple rounds of norm-bounded projected gradient descents to strengthen the backdoor effect. Unlike centralized backdoor attacks on FL where each adversarial participant embeds the same global trigger during training, distributed backdoor attack (DBA) \cite{5} decomposes a global trigger pattern into disjoint local patterns and embeds each of them into the training set of a different adversarial participant. Baruch et al. \cite{6baruch2019little} showed that directed small changes to many parameters of a few Byzantine participants is sufficient to circumventing defenses for FL. Their A Little Is Enough (LIE) attack searches for the perturbation range in which the parameters can be changed without being detected even in the i.i.d. settings. By operating within the population variance, non-omniscient attack with access to only data of the rogue participants is feasible. Bhagoji et al. \cite{3} showed that a highly constrained adversary can still succeed in model poisoning by an alternating minimization strategy with distance constraints to avoid the statistical anomaly detection on updated values. This attack is more stealthy than model replacement \cite{1} as it boosts only the malicious component of the update. 
\subsection{Robust Federated Learning}
\subsubsection{Server-side defenses}
Most of the existing defenses are performed by the trusted federated server. The most widely studied approach is the statistics-based defense leveraging specific aggregation rules to remove or mitigate the outliers. Averaging the received gradients is optimal in the absence of Byzantine clients but brittle even if only one malicious cient is present. Non-linear aggregation rules like median \cite{yin2018byzantine}, trimmed-mean, geometric-mean \cite{pillutla2022robust}, krum and bulyan \cite{blanchard2017machine} have been proposed to replace the arithmetic mean in the original FedAvg by computing a reliable vector or choosing one of the values that represents the centre of the benign distribution. FLTrust \cite{37cao2020fltrust} utilizes the root dataset, which is a small clean dataset collected manually by the server, to bootstrap trust on the server. FLTrust normalizes local model updates to the same hyper-sphere as the server model to diminish the effect of adversarial updates.  To strengthen good updates against Byzantine attacks, an adaptive federated averaging (AFA) algorithm \cite{8munoz2019byzantine} was proposed.  It uses Hidden Markov model to model the ability of each client and cosine similarity to measure the quality of model updates during training. The above defenses deter availability attacks from faulty, noisy and outlier behaviors and block bad clients based on measurable distance between benign and Byzantine model parameters, which may not be effective in detecting backdoor updates. Robust Learning Rate (RLR) method \cite{9ozdayi2021defending} controls the clients’ participation in each dimension of the model updates by adjusting the server's learning rate according to the sign information of the client’s model updates. The idea is to move the aggregated model parameters in the direction towards   maximizing the adversarial loss. Foolsgold \cite{18fung2018mitigating} also makes use of the similarity between malicious clients' gradients to counteract sybil-based label flipping and backdoor poisoning with boundless number of attackers. Instead of identifying suspicious local gradients over plaintexts, ShieldFL \cite{shieldFL22} measures the cosine similarity over encrypted model update using two-trapdoor homomorphic encryption to resist poisoned model without compromising privacy.   

Anomaly detection approaches try to identify anomalous data in the distribution of local model updates and remove them from the aggregation \cite{7li2020learning, rieger2022deepsight}. Several methods have been proposed along this line, for instance, low-dimensional embedding analysis \cite{7li2020learning}, K-means clustering \cite{11shen2016auror}, graph-based anomaly detection, and feedback-based process \cite{12andreina2021baffle}. Defense like \cite{2} clips the L2-norm of the model updates to a fixed value, effectively limiting the contribution of each individual update to the aggregated model.  

These defenses deter the poisoning attacks to some extent but they are not without caveats. Most methods fail to prevent the aggregation from being polluted when aggressive adversaries have control of a large proportion of clients, or the malicious updates are made as close to other updates as possible to bypass the detection. Clipping based methods are ineffective against poisoned model updates with small magnitudes but high attack impact.  Additionally, clipping and detection-based methods can only be operated on plaintext model updates. Many of the robust aggregation rules do not work with privacy-preserving aggregation, which weakens the privacy assurance of FL. Privacy-preserving aggregations like RFA \cite{pillutla2022robust} and ShieldFL \cite{shieldFL22} involving complex cryptographic operations are highly computationally intensive. Similar to other robust aggregation rules, they often reject not only poisoned model updates but also flatten out benign model updates, which cause performance degradation of the federated model. 
 
\subsubsection{Client-side defenses}
Defenses performed at the client's side are more reliable as it provides guarantee for each client individually without assuming the trustworthiness of the server and other clients. Nevertheless, very few methods tackle poisoning attacks from the client perspective. White Blood Cell for FL (FL-WBC) \cite{19sun2021fl} is a client-based defense that mitigates model poisoning effect of polluted global model by perturbing the parameter space during local training where long-lasting attack effect on parameters is identified. Inspired by matching networks, where the class of an input is predicted from the similarity of its features with a support set of labeled examples, Chen et al. \cite{17chen2020backdoor} proposed a new defense mechanism to reduce the impact of backdoor attacks by removing the decision logic from the shared model. Adversarial training can also be leveraged to enhance robustness of FL. Federated adversarial training (FAT)~\cite{20zhang2020defending} enables a learning model to pivot on the sensitive attributes by modifying the optimization objectives and loss functions to achieve predictions that are independent of the sensitive attributes embedded in the training data. 

To reduce the error bound and poisoning susceptibility due to the high dimensionality of model parameter updates, Cronus~\cite{chang2019cronus} abandons the shared model completely by sharing only the logits of locally trained models on public dataset. As adversaries can eavesdrop the transmitted confidence scores corresponding to the public data, black-box access itself is inadequate to avert model extraction and inversion attacks~\cite{tramer2016stealing}. Both membership privacy mechanism for training the local model and prediction privacy mechanism for protecting the updates are necessary but they inevitably trade utility for privacy. This atypical FL setting raises further question on the assessment of performance gain - the contribution due to data from other clients over and above the public dataset is inconsistent among clients. Without a common model architecture, objective inconsistency is exacerbated by the increase in diversity of updates across heterogeneous models unknown to the aggregator. This new FL paradigm introduces more biases and poses a significant challenge to utility-privacy trade-off optimization.

\subsection{Knowledge Distillation (KD)}
KD was first developed to compress a bigger model or an ensemble of well-trained machine learning models into a smaller model without significant quality drop \cite{22hinton2015distilling}. It is also referred to as a teacher-student paradigm for knowledge transfer between a knowledgeable model (teacher) and a learning model (student). The original idea is to let the compact student model mimic the teacher model to achieve comparable or even superior performance. 

Lately, KD has also demonstrated potential benefits in enhancing training performance \cite{24zhang2019your}, adversarial robustness \cite{27papernot2016distillation}, and data augmentation \cite{lee2019rethinking}. According to whether the teacher model is updated simultaneously with the student model, the learning mechanisms of KD can be categorized into three main groups: offline distillation, online distillation, and self-distillation. Most of the previous KD methods work offline, where the teacher model is pre-trained, and only the smaller student model is updated with the guidance of the well-trained teacher model during distillation \cite{22hinton2015distilling}. Online distillation is proposed to eliminate the need 
for a large-capacity high-performance teacher model \cite{34zhang2018deep}. Both the teacher and student models are updated
simultaneously, and the whole KD framework is end-to-end trainable \cite{35chen2020online}. Specifically, mutual learning \cite{34zhang2018deep} and co-distillation \cite{36anil2018large} train multiple neural networks that work with the same architecture in a collaborative way, where any one network can be the student
model and is trained by transferring the knowledge from the
other teacher models. In self-distillation, the same network acts as teacher and student. Zhang et al. \cite{24zhang2019your} first proposed to distill knowledge within the network itself by transferring knowledge from deeper sections of the network into shallow sections. Similarly, self-attention distillation \cite{25hou2019learning} utilizes the attention maps of higher layers as distillation targets for its lower layers. Snapshot distillation \cite{26yang2019snapshot} is
a special variant of self-distillation, in which the supervision from the prior iterations of the network is taken to guide the training of its later iterations. This method requires the difference between teacher and student to be sufficiently large to prevent under-fitting.

Offline KD has been utilized to remove the backdoor of the trained model guided by the finetuned teacher model \cite{li2021neural}. In FL, the defense against backdoor attacks takes place during the training process. The threat model and information available to the defender are completely different from those mentioned above for centralized learning.
 
\section{SPFL: Self-Purified Federated Learning}
\subsection{FL Setting and Threat Model}
\label{sec:FL}
Our defense mechanism focuses on the horizontal FL, where $N$ data owners  collaboratively train a model with $M$ parameters through an aggregator. 
\subsubsection{System Setting}
Let the machine learning model be $f(X, w)$, where $X$ and $w$ are the input and model parameter, respectively. The set of distributed training data of a client $u$ with $n$ samples is denoted as $D_u = \{(x_1, y_1), \cdots,(x_n, y_n)\}$.
The aim of FL is to fit an optimized global model across the distributed local training data of participating clients. The objective can be formulated as:
\begin{equation}
\min_w G(F_1(f_w, D_1),\cdots,F_N(f_w, D_N)) 
\end{equation} 
where $F_k(f_w, D_k)$ denotes the local objective of the $k$-th client, and $G(\cdot)$ denotes the aggregation rule of the local objectives. In FedAvg, $F_k(f_w, D_k) = \frac{1}{n} \sum_{i=1}^{n}L(f_i(x_i, w), y_i)$ is typically an empirical risk minimization by averaging a loss function $L$, such as cross-entropy loss. $G(\cdot)$ is set to be the weighted average of the inputs, i.e., $\frac{\sum_{i=1}^{N} \alpha_i F_i(f_w, D_i)}{\sum_{i=1}^{N} \alpha_i}$. 

A  typical federated process for an iteration $t$ \cite{45} is delineated as follows:
(1) The server broadcasts the aggregated model $w_0^{t-1}$ of the last iteration $t-1$ to the clients;
(2) The client $u$ performs local training for $E$ internal epochs on the local training data $D_u$, starting from $w_{u,0}^t = w_0^{t-1}$. The local model after trained by this iteration is $w_{u}^{t}$;
(3) The server computes the aggregation over the received model update\footnote{The term update and parameter are used interchangeably for the value submitted by a client to the server. The value, often a vector or a tensor that focuses on the main learning task, is transmitted in plain, perturbed or encoded form depending on the aggregation algorithm, which can have a variety of incarnations with different privacy guarantees for different threat models.} and updates the global model by $w_0^t=f_{agg}(w_i^{t})$ for $ i\in[1,N]$. In FedAvg, $w_0^{t} = \frac{\sum{\alpha_{i}w_i^{t}}}{\sum{\alpha_i}}$.
This process is repeated multiple iterations until convergence to obtain the final global model parameters $w_{\textrm{final}}$. 

If $f_{agg}(w_i^{t})$ is computed by secure aggregation, the protected model data $[[w_u^t]]$ instead of the plain $w_u^t$ is exchanged. Thus, each local model update $w_u^t$ can be accessed only by its data owner $u$. The aggregated intermediate model $f_{agg}(w_i^{t})$ and the final trained model $w_{\textrm{final}}$ will be released to the clients. For brevity, we omit the $[[\cdot]]$ symbol of $[[w_u^t]]$ with the understanding that protected updates are allowed. 

\subsubsection{Threat Model}
\label{threat}
We consider two types of threat model in FL \cite{47back}. One threat model assumes that the adversary can have access to only the training data but not the training process. The adversary may be the data provider, who can pollute the training data to create an attack. Another  assumes that the adversary has access to the local model as well as the training data. In addition to modifying the training data, the adversary can also manipulate the training process or updated parameters. We further assume that a determined adversary is powerful enough to fully control a set of up to $N_m$ clients, and may also control the server. The controlled malicious clients could arbitrarily deviate from the protocol. Otherwise, the adversary is assumed to have no control over the rest of the benign clients nor has access to their data or model updates. We also assume that the adversary has full knowledge of the aggregation algorithm and configuration parameters but does not have the ability to tamper with them. 

The goal of the adversary is to compromise the integrity of the trained model, by making the final global model fail only on a specific task while performing reasonably well on the original tasks. In addition, no client can notice that the distributed global model has been compromised even by subjecting the model to a rigorous evaluation of their own dataset. To this end, the controlled client $v$ uses a malicious objective function $H_v(w^{t}, D_v^*)$ to update the local model. Note that $H_v(\cdot)$ is not modified in data poisoning. It merely incorporates the attack goal in model poisoning. It is $D_v^*$ that manipulates the training data. In other words, our defense imposes no constraint on the attack methods used to realize $H_v(\cdot)$ and  $D_v^*$. Any feasible combinations of data and/or model poisoning attack methods can be applied.
\subsubsection{Desiderata of Robust FL}
Based on the above threat model, an active defense for robust FL scheme should possess the following desirable properties:
(1) Integrity security: the scheme should be able to mitigate the influence of poisoning attacks from a large number of malicious clients. The final deployed model should perform normally on the adversary-chosen inputs. 
(2) Model quality: a strong security guarantee should not be achieved at the cost of model quality. The main task performance of the resulting model must be comparable to that without the defense.
(3) No violation of the privacy principle: the method should work with the secure aggregation mechanism of FL. It should not leak or mandate sharing of individual model updates in plaintext. 
(4) No extra assumption: the defense should be implementable without requiring additional assumptions like the existence of a trusted server or public clean dataset that may not be viable in real-world scenarios. 
 
\subsection{Proposed Robust FL Scheme by Self-Distillation}
In what follows, we provide a high-level overview of SPFL before introducing its underlying local training process and ensemble prediction method.  
\subsubsection{Overview of SPFL}
\begin{figure}[htbp]
	\centering
	\includegraphics[width=1\linewidth]{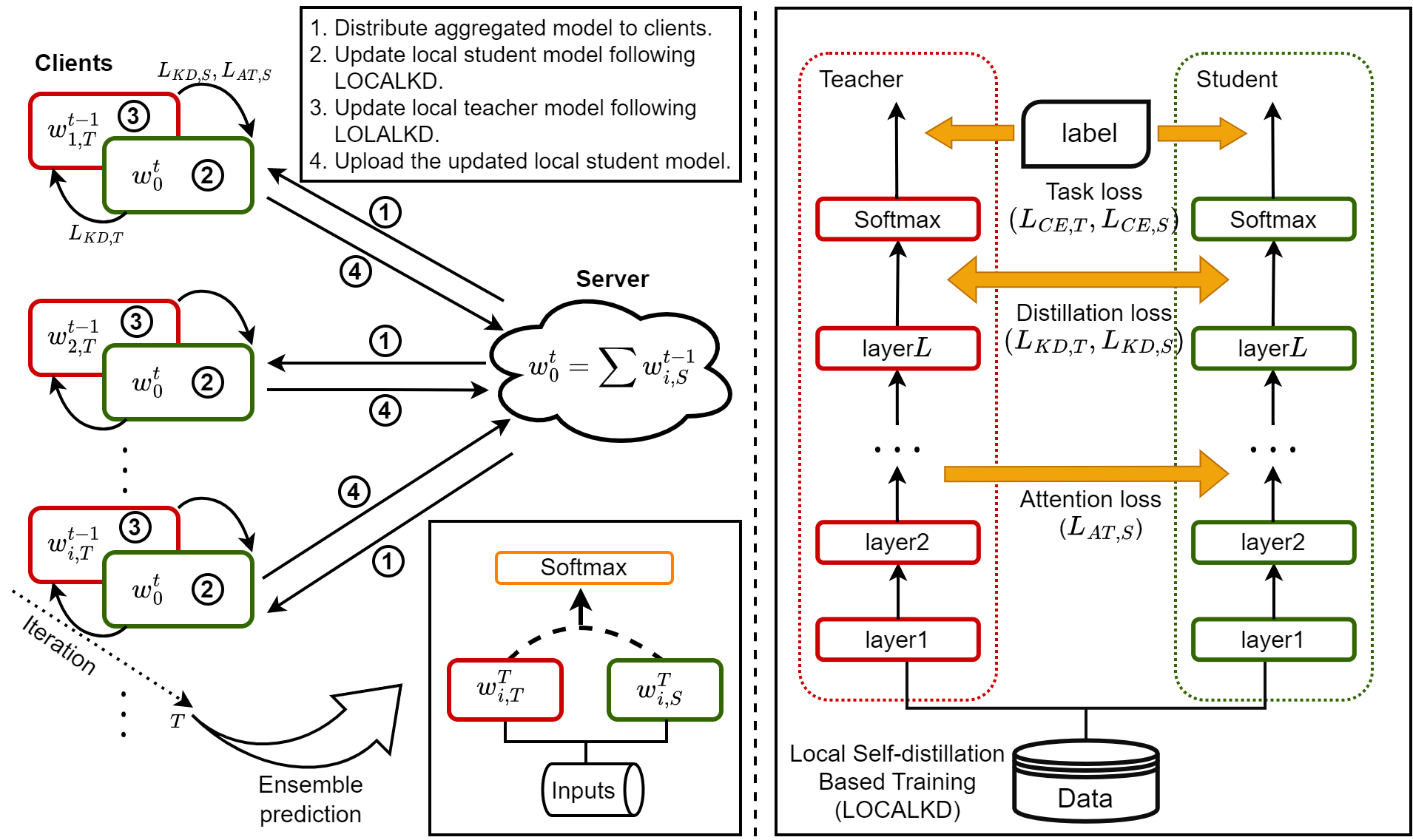}
	\captionsetup{font=small}\caption{An overview of SPFL.}
	\label{overview}
\end{figure}

We illustrate the overall process of SPFL in Fig~\ref{overview}. Our proposed SPFL relies on self-distillation to purify malicious model received by each client. It enables benign clients to take advantage of the historical information of trusted local models to supervise their training process by knowledge transfer. The local teacher and student models are optimized simultaneously during training. At the inference phase, the information of both models are combined to make the final prediction. 

Algorithm 1 depicts the simple modifications SPFL made on the client’s training process. At the beginning of the training, i.e., when $t = 0$, the local (student) model is set as the initialized global model. Each client first trains its student model from scratch independently. In this step, the models are trained by minimizing the task-specific loss function that measures the error between the predicted and ground-truth labels of the local training data. The training is performed iteratively for $E$ local internal epochs. At the end of this step, the vector $w_{i,S}^{0}$ of the trained student model is sent to the server for aggregation. Meanwhile, each client utilize the parameters of locally trained $w_{i,S}^{0}$ to initialize the local teacher model. For each subsequent step $t > 0$ after model aggregation, each client updates the local model following the LOCALKD algorithm (see Algorithm~\ref{alg:B} in Sec.~\ref{sec:localkd}), starting from the aggregated result of the updated model of the last iteration,  which is $w_0^{t} \longleftarrow f_{agg}(w_{i,S}^{t}) \forall i\in[1, N]$. Self-KD instead of SGD is performed by the benign clients to update the local model in LOCALKD algorithm. The aim is to mitigate possible pollutions introduced into the aggregated result by the adversaries. The teacher model $w_{i,T}^{t-1}$ is the uncontaminated local model of the last iteration before aggregation. The student model $w_0^t$ is the aggregated model to be trained. Both are updated in reciprocity to enhance the integrity and performance of the trained model. Then, the updated teacher model is used for teaching in the next iteration, and the parameters of updated student model are uploaded to the server for aggregation. This process is repeated until the model converges, or the preset number of training iterations has been reached. Through this process, the student model learns not only from local data during the LOCALKD training, but also from data of other clients during the aggregation. If the global model is corrupted by updates from malicious clients through aggregation, the local teacher model helps to correct the malicious behavior via distillation. Meantime, the teacher model also benefits from the maximization of mutual information by knowledge distilled from the student model. It should be noted that $f_{agg}(\cdot)$ and $[[\cdot]]$ in Algorithm~\ref{alg:A} are generic, which allows the purification to be made regardless of the FL aggregation algorithms or privacy-preserving and security measures implemented at the server.

After completing the training process, each client obtains a personalized teacher model and a personalized student model. To make the prediction more robust, each client runs ENSPRE algorithm (see Sec~\ref{sec:ENPRE}), which merges the features  of both models to output the prediction of the input sample. 

\begin{algorithm} \small
\caption{Robust Federated Learning Scheme}  
\label{alg:A} 
\begin{algorithmic}
\STATE Input: Initialized global model $w_0$\\
\STATE \underline {During training phase:}
\FOR {iteration $t\in [0,T]$ }
    \IF{$t=0$}
    \FOR{ $i\in[1,N]$ clients in parallel}
    
      \STATE Initialize the local student model $w_{i,S}^t$ as $w_0$\\
      \STATE Update local student model $w_{i,S}^t$ with FedAvg steps\\
      \STATE Initialize the local teacher model $w_{i,T}^{t}$ as $w_{i,S}^t$ \\
    
      \STATE Send $[[w_{i,S}^{t}]]$ to the server\\
    
    \ENDFOR
    
    \ELSE
    \FOR{ $i\in[1,N]$ clients in parallel}
    
      \STATE Update $w_{i,S}^{t}, w_{i,T}^{t}\longleftarrow \textrm{LOCALKD}(w_0^t, w_{i,T}^{t-1},D_i)$\\
    
      \STATE Send $[[w_{i,S}^{t}]]$ to the server\\
    
    \ENDFOR
    \ENDIF
   \STATE Server does:\\
   \STATE $w_0^{t+1} \longleftarrow f_{agg}(w_{i,S}^{t}) $ for $ i\in[1,N] $\\
    \ENDFOR
\STATE \underline{During prediction phase:}
\FOR{ $i\in[1, N]$ clients }
    
      \STATE Given input $X$, output $Y\longleftarrow \textrm{ENSPRE}(w_{i,S}^{T}, w_{i,T}^{T},X)$\\
    \ENDFOR
\end{algorithmic}
\end{algorithm}

\subsubsection{Local Self-distillation Based Training}
\label{sec:localkd}
The self-distillation based local training algorithm, LOCALKD is described in Algorithm~\ref{alg:B}. In training iteration $t$, each client $i$ simultaneously updates its teacher and student models with the help of labeled local training data $D_i$. Unlike traditional KD, the teacher model $w_{i,T}^{t-1}$ and student model $w_0^t$ are essentially the same model from different iterations. The teacher model is kept as a local model and is trained by only the local training data with self-KD, but the student model is re-instantiated by the global model of the last aggregation at the server. 

The local training algorithm needs to fulfil two objectives. First, the teacher model is optimized for main task accuracy. To facilitate learning from the local training data and knowledge distillation from the student model, we divide the loss function into task loss, which is used to provide direct task-specific supervision, and distillation loss, which is used to transfer knowledge from the logit output. Second, the student model must be able to mitigate the backdoor effect by knowledge transfer from the teacher model besides improving its main task performance. From the explainability study of backdoor attacks, differences in the saliency maps of intermediate layers between the backdoored and clean models provide useful clues to backdoor trigger localization~\cite{48}. To suppress the backdoor triggers if presence, a third loss function that minimizes the aggregated difference between the post-hoc attention maps of the student and teacher models is added to train the student model.

The training is performed locally by sampling mini-batches or subsets from the training set. Each sampled set, $B_l$, is fed into both the teacher and student models. For an input $x$ and a $k$-dimensional one-hot label $y$, a model with parameter $w$ produces the logit vectors $z(w, x)=[z_1(x), z_2(x),...,z_k(x)]$, and outputs the predicted probabilities $P(w, x) = [p_1(x), p_2(x),...,p_k(x)]$ by a softmax function. Given the training data $X$ and its corresponding label $Y$ in $B_l$, the task loss function estimates the difference between the predicted and ground-truth labels by:
\begin{equation}
    L_{CE,S} = \phi(Y, P(w_{i,S}^t, X)) \label{eqa:1}
\end{equation}
\begin{equation}
 L_{CE, T}= \phi(Y, P(w_{i,T}^t, X)) \label{eqa:2}    
\end{equation}
where $\phi(a, b) = -\sum_{i}a_i\log b_i$ is the cross-entropy loss. This loss is minimized over the parameter space to find the optimal parameters that best fits $D_i$.

The distillation losses for the student and teacher models are formulated based on the Kullback–Leibler (K-L) divergence, i.e., $KL(a, b) = -\sum_{i} a_i \log( b_i/a_i)$, between the softened probabilities of the teacher and student models, as follows:
\begin{equation}\label{eqa:4}
    L_{KD,S} = KL(\widetilde{P(w_{i,T}^t, X)}, \widetilde{P(w_{i,S}^t, X)})
\end{equation}
\begin{equation}\label{eqa:5}
   L_{KD,T} = KL(\widetilde{P(w_{i,S}^t, X)}, \widetilde{P(w_{i,T}^t, X)})
\end{equation}
where the softened probability $\widetilde{P(w, x)}$ is generated by the logit vectors with $\tau$ as the temperature parameter to adjust the degree of softening as follows~\cite{22hinton2015distilling}:
\begin{equation}\label{eqa:3}
  \widetilde{P(w, x)} = \frac{exp(z_i(X)/\tau)}{\sum_j exp(z_j(X)/\tau }  
\end{equation}

The attention-based loss is measured by the distance between the attention maps of the two models. Given the activation output of the $l$-th layer, $A^l \in \mathbb{R}^{K\times U \times V}$, where $K, U$ and $V$ are the dimensions of the channel, height and width of the feature map, respectively, the class-discriminative attention representation $L^c\in \mathbb{R}^{U\times V}$ for class $c$ is computed by:
\begin{equation}\label{eqa:6}
    L^c(A^l) = RELU(\sum_{k=1}^K \alpha_k^c A_k)
\end{equation}
where $A_k$ is the activation map of the $k$-th channel.The weight $\alpha_k^c$ represents a partial linearization of the deep network downstream from $A$. Backpropagation is employed to compute the gradient of the score for the target class $c$ before the softmax function, with respect to the activation map $A_k$ of a convolutional layer. Subsequently, this gradient information is utilized to assess the significance of each channel within feature layer $A$ by performing a weighted summation on the data in each channel of feature layer $A$ using the values of $\alpha_k^c$.  Finally, a Rectified Linear Unit (ReLU) activation function is applied to the linear combination of maps to emphasize the features that exert a positive influence on the class of interest.

Based on the attention representation, an attention loss is defined for the selected intermediate layer $l\in L$ as:
\begin{equation}\label{eqa:7}
   L_{AT, S} = \sum_l \left \|L^c (A_T^l) - L^c (A_S^l)) \right \|_2
\end{equation}
where $L_2$ norm is used to measure the distance between the teacher's attention map $L^c (A_T^l)$ and the student's attention map $L^c (A_S^l)$.

Bringing it all together, the overall training losses for the student and teacher models can be expressed as:
\begin{equation}\label{eqa:8}
   L_S = L_{CE,S} + \beta_{KD}^t L_{KD, S}+ \beta_{AT}^t L_{AT, S}
\end{equation}
\begin{equation}\label{eqa:9}
  L_T = L_{CE,T} + \beta_{KD}^t L_{KD, T}
\end{equation}
where $\beta_{KD}^t$ and $\beta_{AT}^t$ are adaptive scaling factors for controlling the strength of KD between the two models.  

During the distillation process, the incorrect predictions from the teacher or student model may mislead each other in the knowledge transfer. It is important to adapt the distillation effort to the quality of trained model by changing the weight of distillation loss progressively along the training process. The model generally does not have enough knowledge about data at the early training stage, and its generalization capability improves gradually as the training progresses. So the adaptation can be simplified by increasing the values of $\beta_{KD}^t$ and $\beta_{AT}^t$ gradually with the training iteration $t$. We compute $\beta_{KD}^t$ and $\beta_{AT}^t$ for each iteration using a linear growth rate, i.e.,  
\begin{equation}\label{eqa:linear}
  \beta_{\cdot}^t = \beta_{\cdot}^T \times \frac{t}{T}
\end{equation}
where $T$ is the total number of training iterations and $\beta_{\cdot}^T $ is the weight at the last iteration. 

Each optimization is performed in two phases: (i) update the student model by minimizing the individual loss via gradient descent. (ii) update the teacher model whose distillation loss is computed based on the updated student model. This process is iteratively performed for $E$ internal epochs to obtain the trained local student model $w_{i,S }^{t}$ and teacher model $w_{i,T }^{t}$. Parameters of $w_{i,S }^{t}$ are sent to the server, while the trained teacher model $w_{i,T}^t$ is retained locally to guide the next iteration of training.
\begin{algorithm}
\caption{Local Self-distillation Based Training}  
\label{alg:B} 
\begin{algorithmic}\small
\STATE Input: individual training data $D_i$, number of internal training epochs $E$, training configurations, global model $w_0^t$ \\
\STATE Initialize local student model $w_{i,S}^t = w_0^t$, prepare the local teacher model  $w_{i,T}^t = W_{i,T}^{t-1}$
\FOR{internal epoch $e = 1,2,...E$ }
\FOR{mini-batch $B_l=[X, Y]$ in $D_i$ }
   \STATE Compute task loss $L_{CE, S}, L_{CE, T}$ by (\ref{eqa:1}) and (\ref{eqa:2})
   \STATE Compute distillation loss for student $L_{KD, S}, L_{AT, S}$ by (\ref{eqa:4}) and (\ref{eqa:7})
   \STATE Update student model by $w_{i,S}^t = w_{i,S}^t - \gamma \bigtriangledown_{w_{i,S}^t}(L_{CE,S} + \beta_{KD}^t L_{KD,S} + \beta_{AT}^t L_{AT, S})$
   \STATE Compute distillation loss for teacher $L_{KD, T}$ by  (\ref{eqa:5})
   \STATE Update teacher model by $w_{i,T}^{t} = w_{i,T}^t - \gamma \bigtriangledown_{w_{i,T}^t}(L_{CE,T} + \beta_{KD}^t L_{KD,T})$
   
\ENDFOR
    
    \ENDFOR
\RETURN{$w_{i,T}^{t}, w_{i,S}^{t}$ }
\end{algorithmic}
\end{algorithm}
\subsubsection{Ensemble Prediction Method}
\label{sec:ENPRE}
The logit outputs of the trained teacher and student models are averaged to further neutralize any malicious residues from the student model. For client $i$, given an input sample $X$, the algorithm $\textrm{ENSPRE}(w_{i,S}^{T}, w_{i,T}^{T},X)$ first computes the ensemble logits by
\begin{equation}\label{eqa:11}
   z(w_{i}, X) = \frac{z(w_{i,T}, X)+z(w_{i,S}, X)}{2}
\end{equation}

Then the softmax function is operated on the averaging logits $ z(w_{i}, X)$ to generate the final classification result. 
\section{Experimental Results and Discussions}
\subsection{Experimental Setting}
\subsubsection{FL System Setting} Two representative machine learning applications are trained in the FL setting on which most of the attack methods can be applied for comparison. The first application is a handwritten digit recognition task. The CNN model with two convolutional layers and about 0.2M parameters is trained on the MNIST dataset~\cite{mnist}. The second application is an image recognition task.  ResNet18 \cite{resnet} with 11M parameters is trained on the CIFAR10 dataset \cite{2012Learning}.
Six aggregation methods are used for each application: (i) FedAvg~\cite{40kairouz2021advances}, the most basic aggregation without any defense; (ii) Median aggregation~\cite{yin2018byzantine}, an element-wise aggregation rule. For the $j$-th model parameter, the master device sorts the $j$-th parameters of all the local models and takes the median as the $j$-th parameter of the global model; (iii) Approximate geometric mean aggregation (RFA)~\cite{pillutla2022robust}, a nonlinear aggregation method that can detect more nuanced outliers beyond the worst-case malicious setting, and has enhanced security over other robust aggregation rules. It is implemented by calling a regular secure average oracle with built-in privacy guarantees; (iv) FL-WBC~\cite{19sun2021fl}, a client-side defense where the clients perturb the parameter space impacted by long-lasting attack during local training to mitigate model poisoning attacks; (v) RLR~\cite{9ozdayi2021defending}, a server’s side backdoor defense method. For every dimension where the sum of signs of updates is less than the selected learning threshold $\theta$, the aggregation server’s learning rate is multiplied by $-$1 to maximize the loss on that dimension; and (vi) SPFL with average aggregation. For each of these aggregation settings, $T$ rounds of FL are performed among $N=10$ clients. We set $T=30$ for MNIST, and $T=80$ for CIFAR10. 
\subsubsection{Poisoning Attacks and Configurations} Based on the threat model presented in Sec~\ref{sec:FL}, four state-of-the-art backdoor attack methods with various configurations are applied: (i) DPA~\cite{1}. The attacker simply trains its model on backdoored inputs. Each training batch consists of correctly labeled ground-truth and mislabeled backdoor samples \cite{gu2017badnets}; (ii) MPA~\cite{1}. The attack is mounted by training the malicious model $X$ with poisoned training data, and then scaling up its weights to ensure that its submitted contribution survives averaging. The submitted update is calculated as $w_m = \gamma(X-w_0^t)+w_0^t$; (iii) DBA~\cite{5}. This attack involves several coordinated adversarial clients (sybils), where all sybils only use parts of the global trigger to poison their local models; and (iv) LIE~\cite{6baruch2019little}. This method finds a set of parameters within the specific range that the backdoor can be introduced without being detected, while having minimum accuracy loss for benign inputs. 
 
An attack succeeds if the trained model can misclassify the triggered examples to the target label. To increase the attack success rate, during the attacker’s training, we add a special pixel pattern to 20 images in a batch of 64 \cite{1}, and change their labels to 0 for MNIST and bird for CIFAR10. Especially for DBA, the global trigger pattern is divided into multiple local triggers and each local trigger is added by one of the colluding adversaries. At the inference time, the inputs are added with the same pixel pattern to trigger the backdoor. We adjust the configuration of each attack method to change the attack strength. The configurable factors are: (i) Attack frequency ($F_m$): For DPA and LIE, the attack can be made more effective by repeating it at every iteration of FL. For MPA and DBA, multiple-shot (MS) and single-shot (SS) attacks are possible \cite{5}; (ii) Scaling factor ($\gamma$): For single-shot  MPA and DBA, $\gamma$ is set to 10. This is determined by $\gamma=N/\eta$ \cite{1} to guarantee that the backdoor survives averaging when the global model is replaced by the backdoored model\footnote{$N$ represents the total number of clients, and $\eta$ denotes global learning rate which controls the fraction of joint models that are updated every round. Our setting is based on $n=10$ and $\eta=1$.}. For multiple-shot MPA, $\gamma$ is set to 5, which guarantees the attack effectiveness and utility of the global model. We found out experimentally that when $\gamma$ is increased to 10, the main task accuracy of the aggregated model will be severely degraded. For multiple-shot DBA, we set $\gamma=1$ to support the every-round attack of the original paper; (iii) Number of adversaries at each attacking iteration ($N_m$): For DPA and LIE, we evaluate $N_m = 5$ and $N_m=9$ out of 10 clients. For DBA, $N_m$ is set to 4 and 6 out of 10 clients as in~\cite{5}. For MPA, if multiple adversaries exist in a given round, their respective scaled-up updates are divided evenly to keep the scaling valid. As the attack strength is independent of $N_m$, $N_m$ is set to 1. In general, state-of-the-art attacks \cite{6baruch2019little, fang2020local} and defense \cite{lyu2022privacy} assume up to 50\% of FL clients can be compromised. Hence, we test our scheme with up to at least this bound. The extreme attack with only one benign client is also tested. The abbreviations of the canonical attacks and their characteristics are listed in Table~\ref{tab:data}.
\begin{table}
    \centering
    \captionsetup{font=small}\caption{Canonical attacks used in our evaluation.}\label{tab:data}
    \begin{tabular}{rllll}
      \toprule 
      \bfseries Attack & \bfseries Method & \bfseries $N_m$ &  \bfseries $F_m$& \bfseries $\gamma$\\
      \midrule 
      DPA-5 & DPA & 5& every iteration & 1\\
      DPA-9 & DPA & 9& every iteration & 1\\
      LIE-5 & LIE & 5 & every iteration & 1\\
      LIE-9 & LIE & 9 & every iteration & 1\\
      MPA-SS & MPA & 1 & one-shot & 10\\
      MPA-MS & MPA & 1& every 5 iterations& 5\\
      DBA-4*SS & DBA & 4& one-shot & 10\\
      DBA-4*MS & DBA & 4& every iteration & 1\\
      DBA-6*SS & DBA & 6& one-shot & 10\\
      DBA-6*MS & DBA & 6& every iteration & 1\\
      
      \bottomrule 
    \end{tabular}
\end{table}
\subsubsection{Key Performance Metrics}  Three metrics are used for performance evaluation. The \textbf{benign accuracy (BA)} refers to the prediction accuracy on clean samples of personalized models when all the clients are benign. The \textbf{main task accuracy (MA)} denotes the prediction accuracy of a personalized model on clean samples under attack. The \textbf{attack success rate (ASR)} is the ratio of backdoored examples that are misclassified as the target label. Considering that different personalized models are obtained by each client, the average value of all the client models is used to measure each metric. An effective defense should reduce the ASR, while attaining a high MA when the training is corrupted by poisoning attacks. BA is the baseline with no attack. A practical defense should not degrade BA.
\subsection{Results and Comparison}
\begin{figure*}[htbp]
	\centering
	\begin{subfigure}{0.49\linewidth}
	\centering
	\includegraphics[width=1\linewidth]{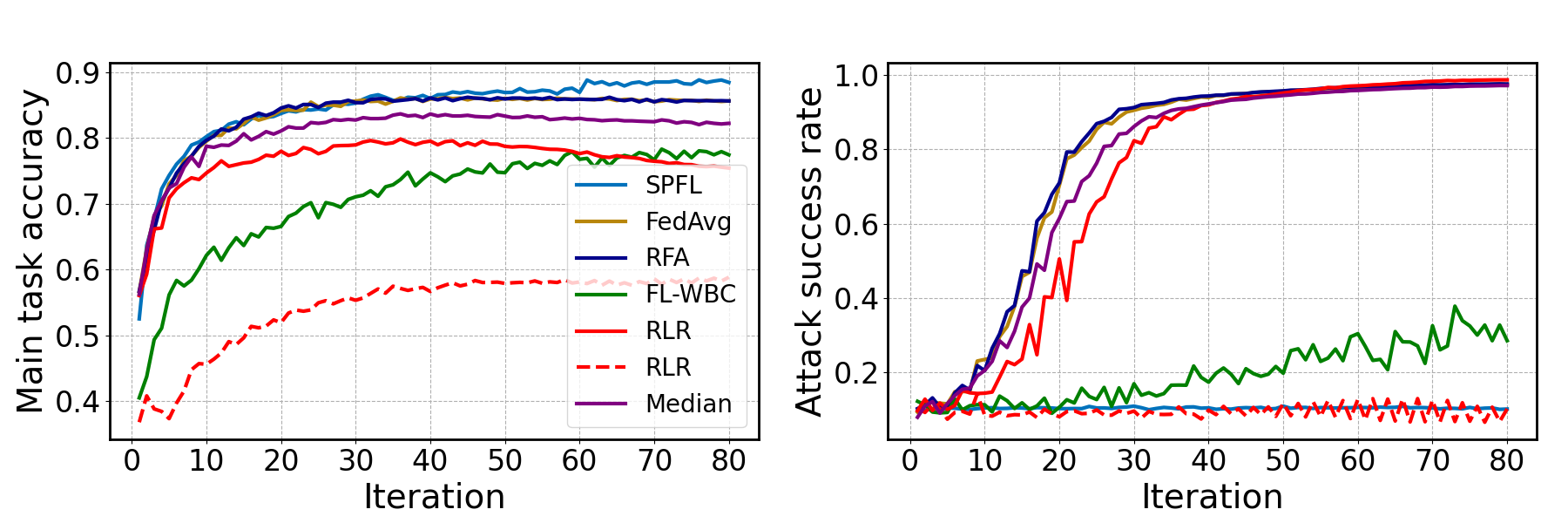}
		\setlength{\belowcaptionskip}{0.1cm} \caption{DPA-5}
	\end{subfigure}
	\centering
	\begin{subfigure}{0.49\linewidth}
		\centering
		\includegraphics[width=1\linewidth]{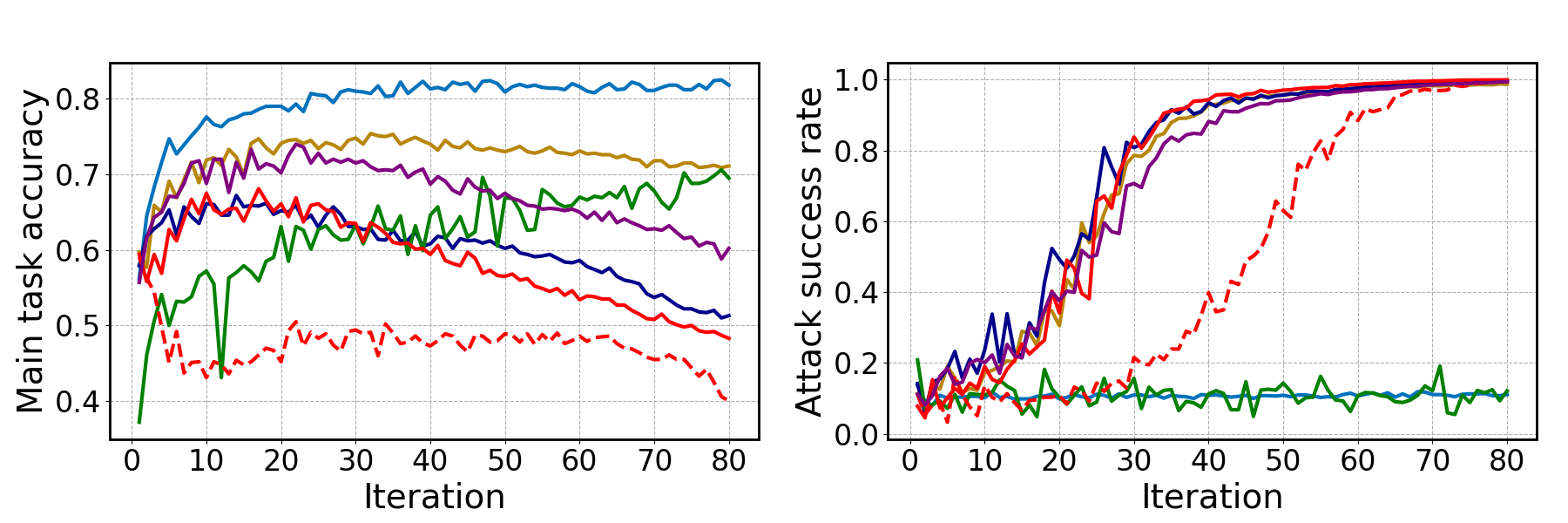}
		\setlength{\belowcaptionskip}{0.1cm} \caption{DPA-9}
	\end{subfigure}
		\centering
	\begin{subfigure}{0.49\linewidth}
		\centering
		\includegraphics[width=1\linewidth]{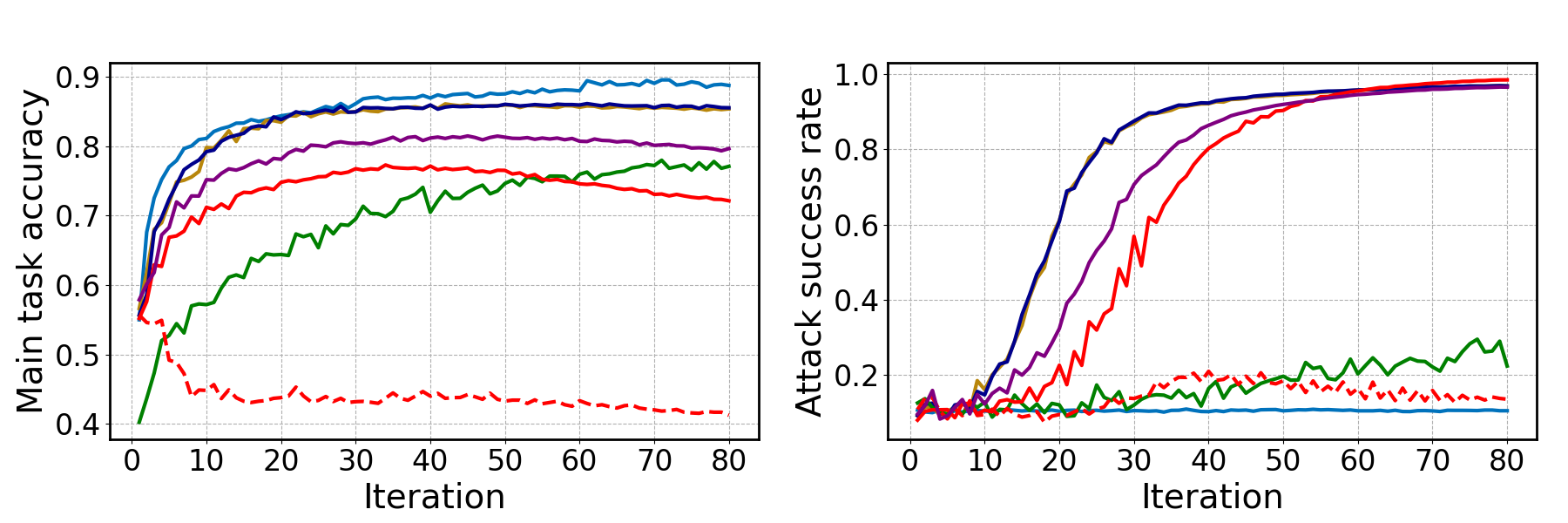 }
		\setlength{\belowcaptionskip}{0.1cm} \caption{LIE-5}
	\end{subfigure}
		\centering
	\begin{subfigure}{0.49\linewidth}
		\centering
		\includegraphics[width=1\linewidth]{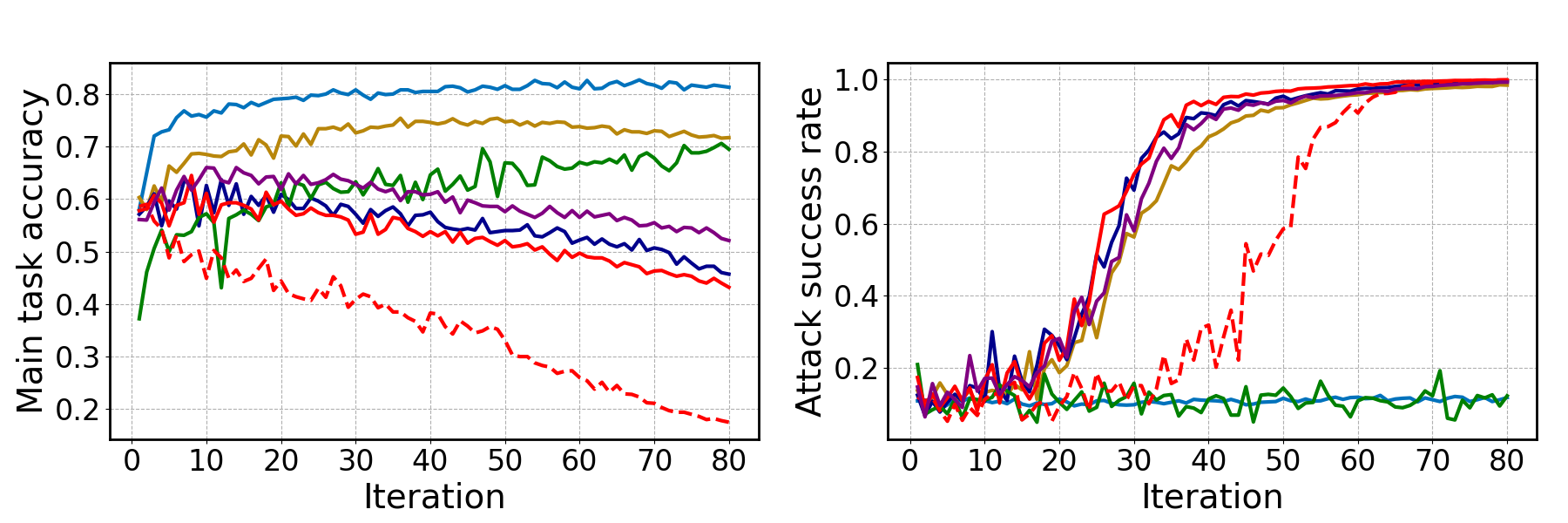}
		\setlength{\belowcaptionskip}{0.1cm} \caption{LIE-9}
	\end{subfigure}
		\centering
	\begin{subfigure}{0.49\linewidth}
		\centering
		\includegraphics[width=1\linewidth]{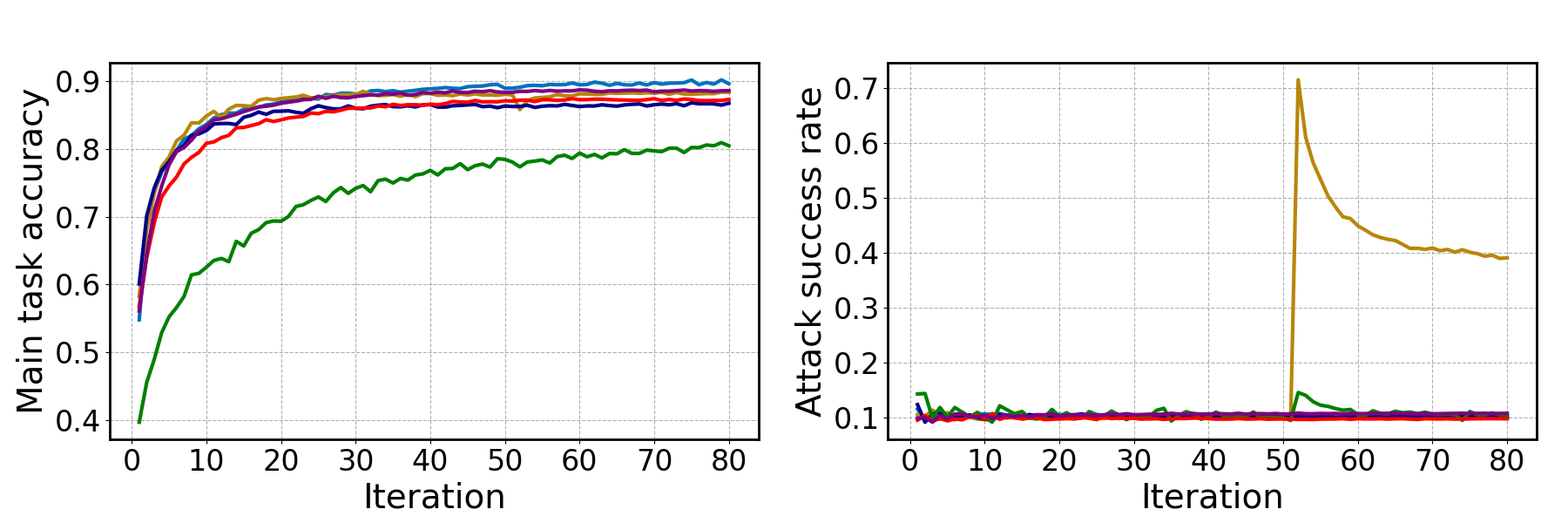}
		\setlength{\belowcaptionskip}{0.1cm} \caption{MPA-SS}
	\end{subfigure}
		\centering
	\begin{subfigure}{0.49\linewidth}
		\centering
		\includegraphics[width=1\linewidth]{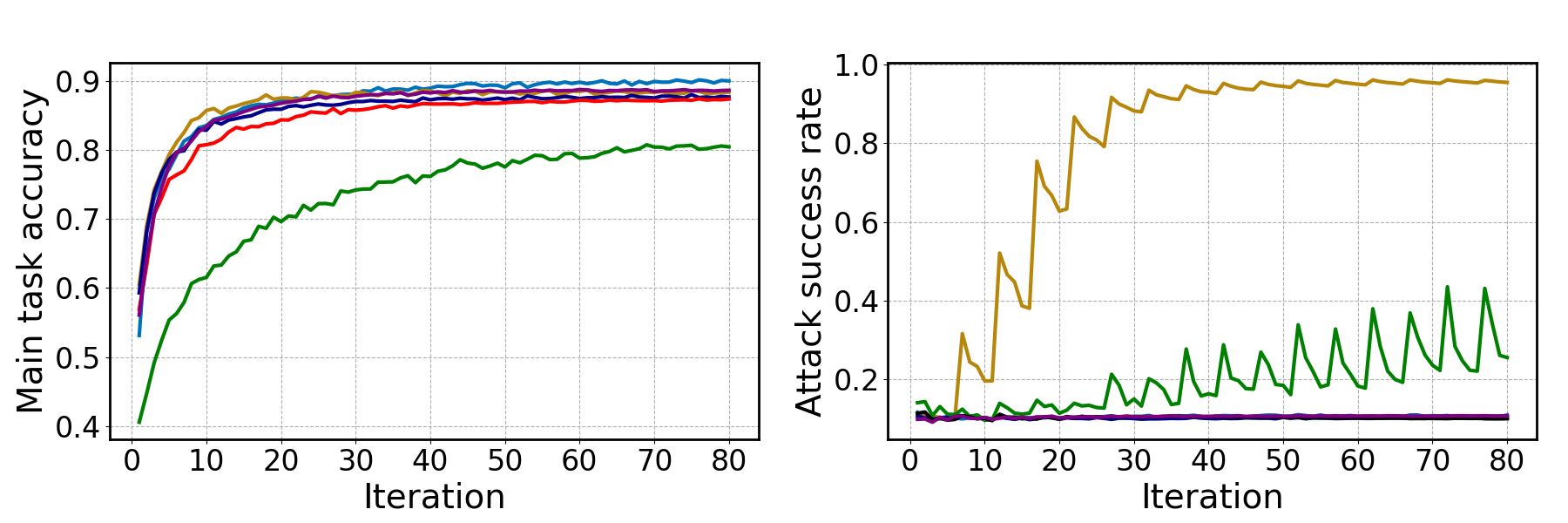}
		\setlength{\belowcaptionskip}{0.1cm} \caption{MPA-MS}
		
	\end{subfigure}
 \centering
	\begin{subfigure}{0.49\linewidth}
		\centering
		\includegraphics[width=1\linewidth]{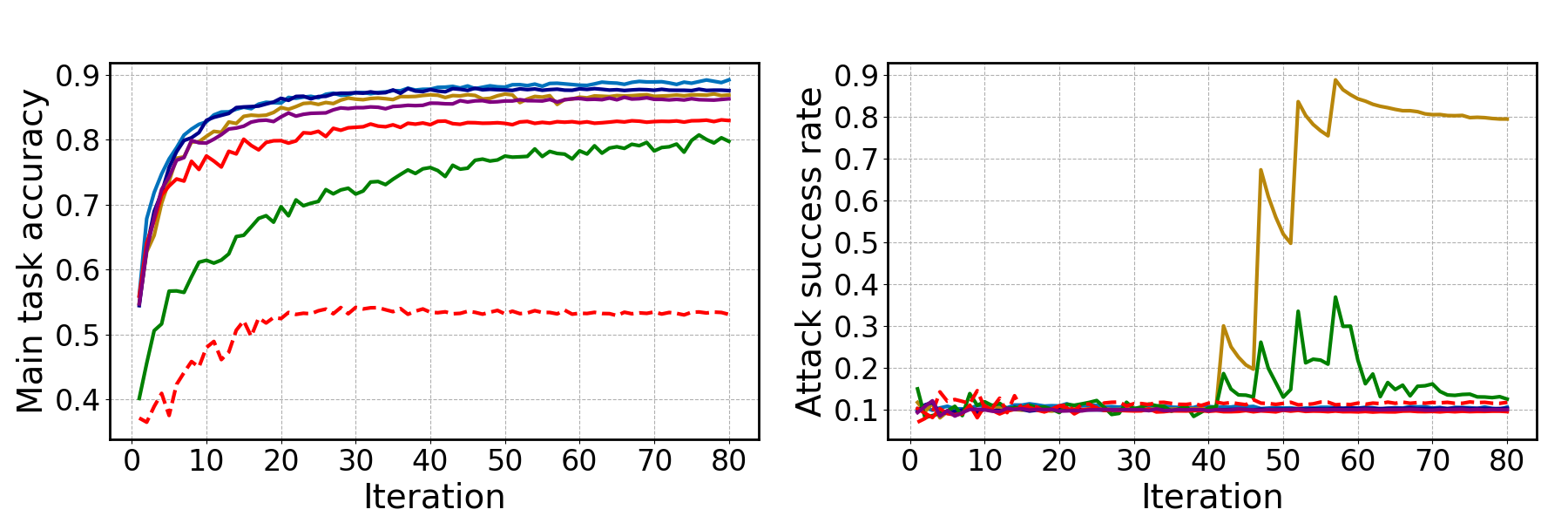}
		\setlength{\belowcaptionskip}{0.1cm} \caption{DBA-4*SS}
		
	\end{subfigure}
 \centering
	\begin{subfigure}{0.49\linewidth}
		\centering
		\includegraphics[width=1\linewidth]{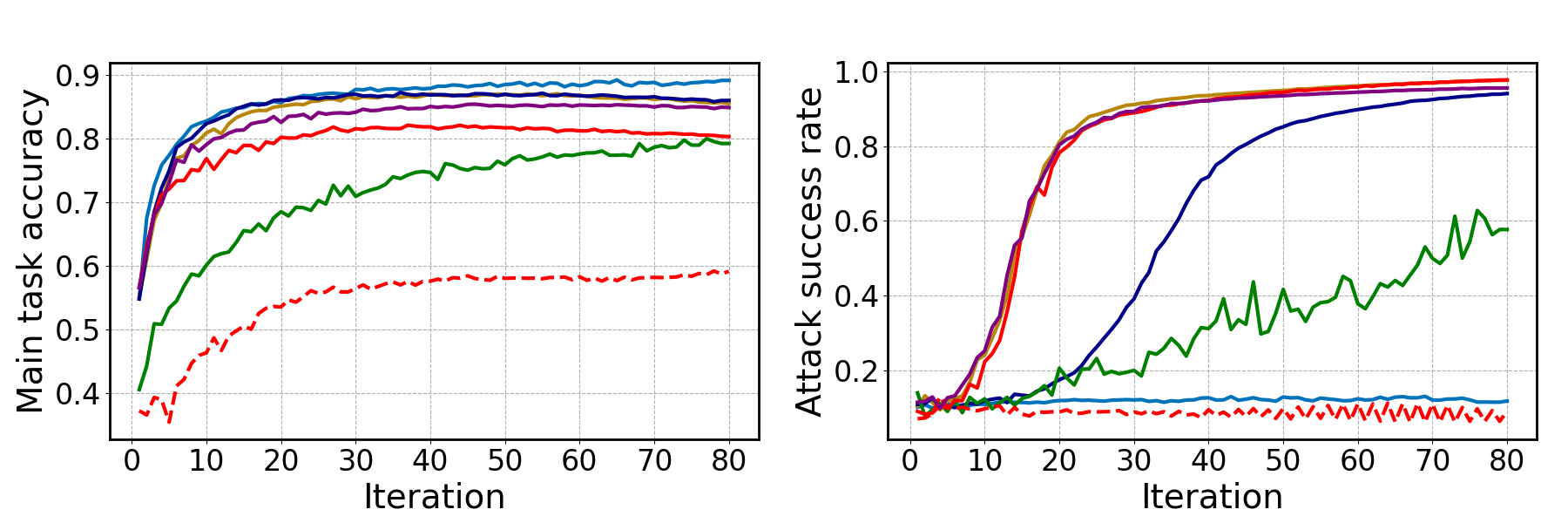}
		\setlength{\belowcaptionskip}{0.1cm} \caption{DBA-4*MS}
		
	\end{subfigure}
 \centering
	\begin{subfigure}{0.49\linewidth}
		\centering
		\includegraphics[width=1\linewidth]{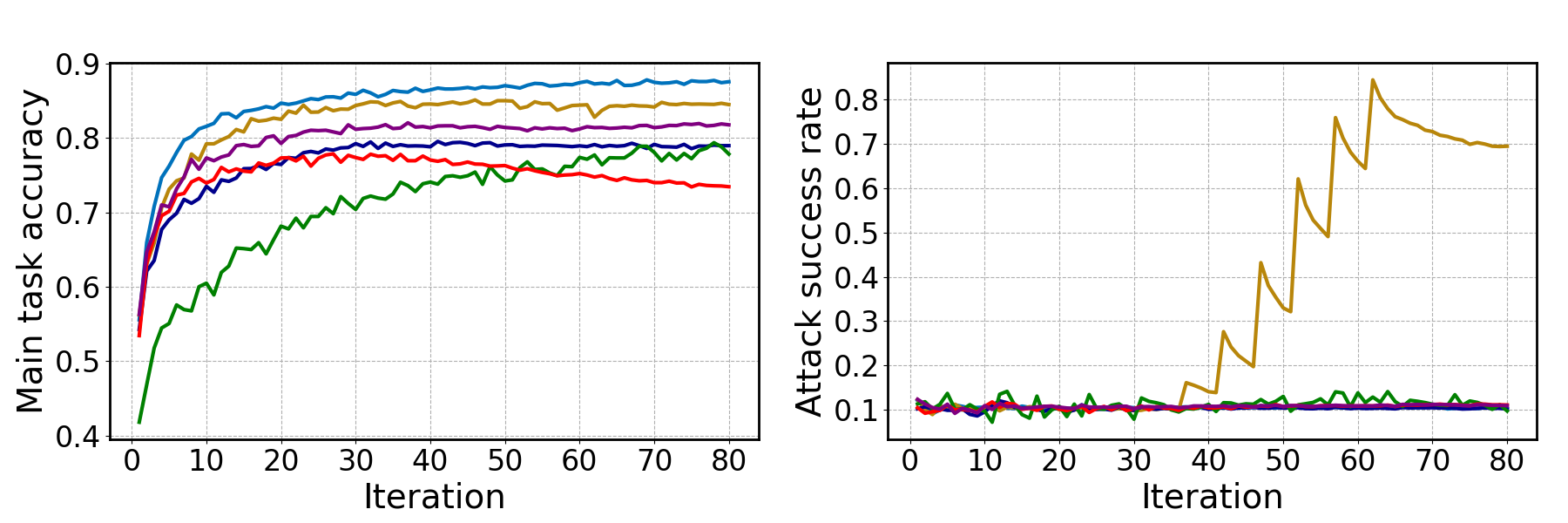}
		\setlength{\belowcaptionskip}{0.1cm} \caption{DBA-6*SS}
		\label{chutian3}
	\end{subfigure}
 \centering
	\begin{subfigure}{0.49\linewidth}
		\centering
		\includegraphics[width=1\linewidth]{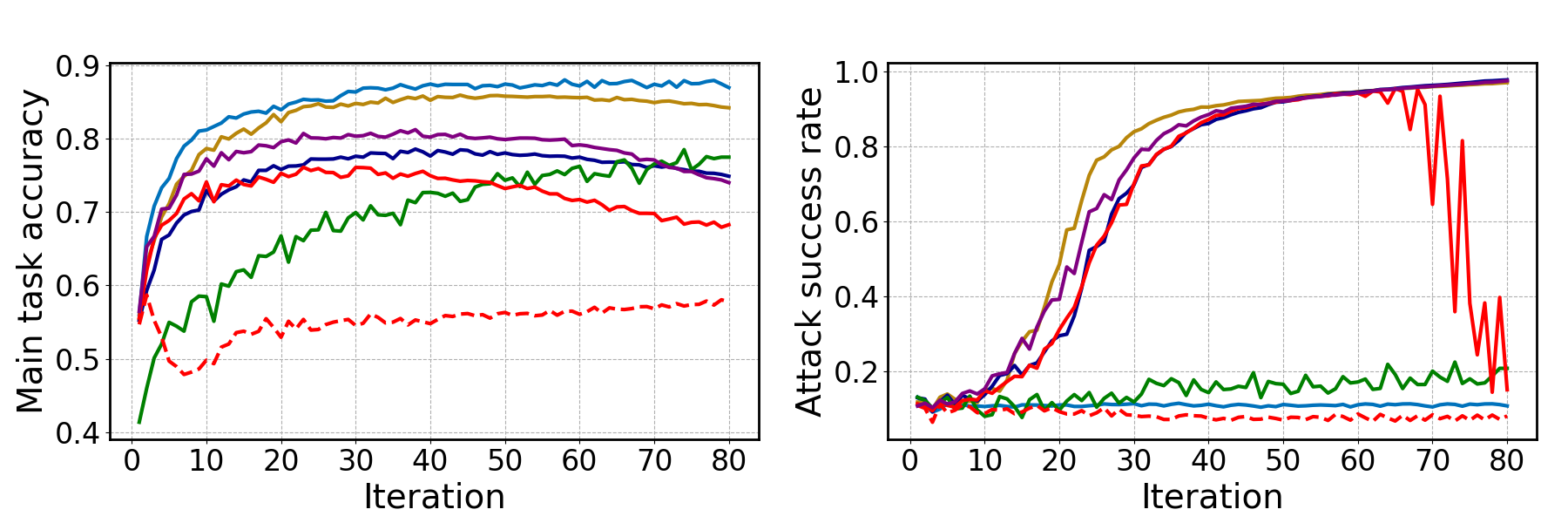}
		\setlength{\belowcaptionskip}{0.1cm} \caption{DBA-6*MS}
		\label{chutian3}
	\end{subfigure}
 
 \captionsetup{font=small}\caption{Main task accuracy (left) and attack success rate (right) for canonical attacks on CIFAR10. Different defense methods are represented by different line colors given by the legend on the top-left subfigure. The red solid and dashed lines refer to RLR with $\theta = 4$ and $6$, respectively. Each subfigure presents the results of one canonical attack: (a) DPA-5 and (b) DPA-9 with attack mounted at every iteration without scaling up the shared malicious model parameters; (c) LIE-5 and (d) LIE-9 with attack mounted at every iteration without scaling up the shared malicious model parameters; (e) MPA-SS with single-shot attack mounted at the 50-th iteration with a scaling factor of 10; (f) MPA-MS with the attack mounted every 5 iterations with a scaling factor of 5; (g) DBA-4 and (i) DBA-6*SS with  one-shot attack mounted at the  41-th (36-th for DBA-6*SS), 46-th, 51-th and 56-th (61-th for DBA-6*SS) iterations with a scaling factor of 10; (h) DBA-4*MS and (j) DBA-6*MS with the attack mounted every iteration without scaling up the shared malicious model parameters.}
	\label{DE_CIFAR}
\end{figure*}

The MA and ASR of each FL method with increasing iterations for MNIST and CIFAR10 datasets are shown in Fig. \ref{DE_MNIST} and Fig. \ref{DE_CIFAR}, respectively. Each subfigure corresponds to one of the attacks listed in Table~\ref{tab:data}. Different line colors represent different defense methods, as indicated by the legend of Fig. \ref{DE_MNIST}(a). From these figures, we can conclude that: 
 
\begin{figure*}[htbp]
	\centering
	\begin{subfigure}{0.49\linewidth}
		\centering
		\includegraphics[width=1\linewidth]{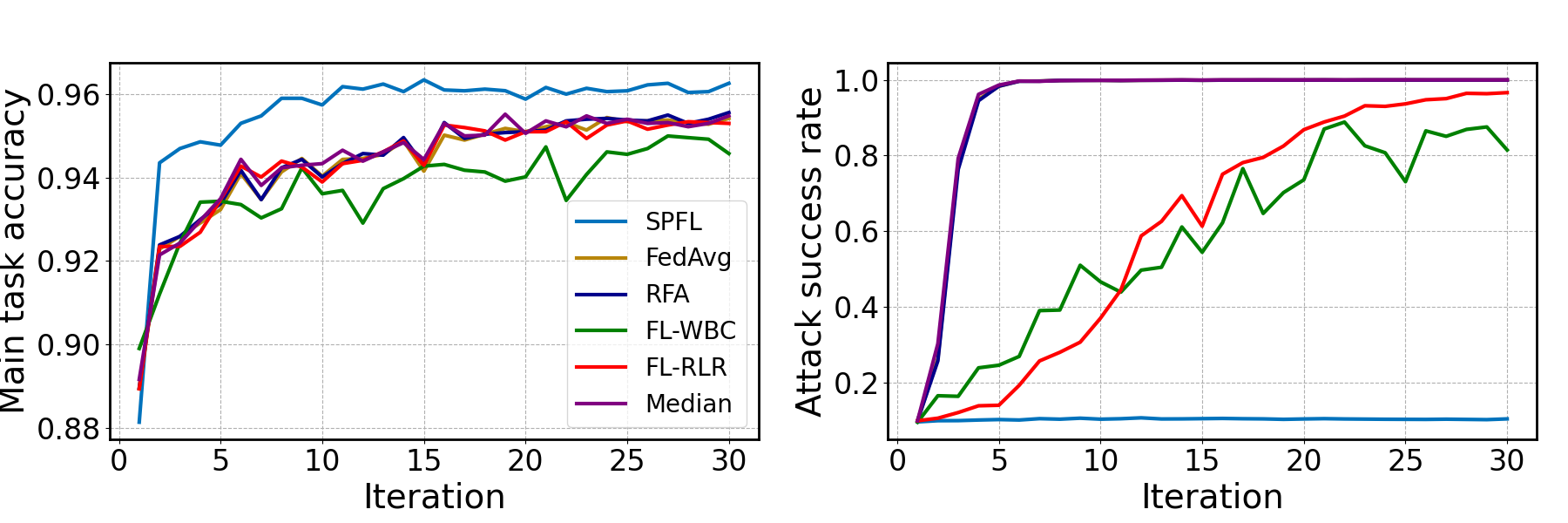}
		\setlength{\belowcaptionskip}{0.1cm} \caption{DPA-5}
		\label{dpa5}
	\end{subfigure}
	\centering
	\begin{subfigure}{0.49\linewidth}
		\centering
		\includegraphics[width=1\linewidth]{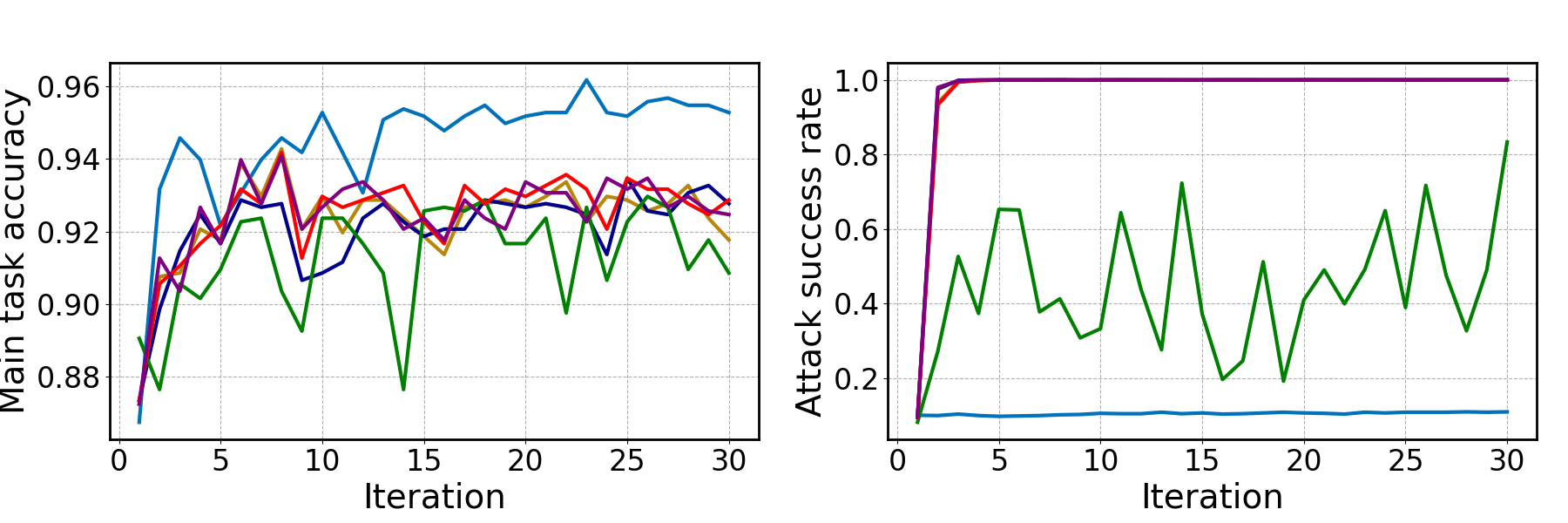}
		\setlength{\belowcaptionskip}{0.1cm} \caption{DPA-9}
		\label{dpa9}
	\end{subfigure}
		\centering
	\begin{subfigure}{0.49\linewidth}
		\centering
		\includegraphics[width=1\linewidth]{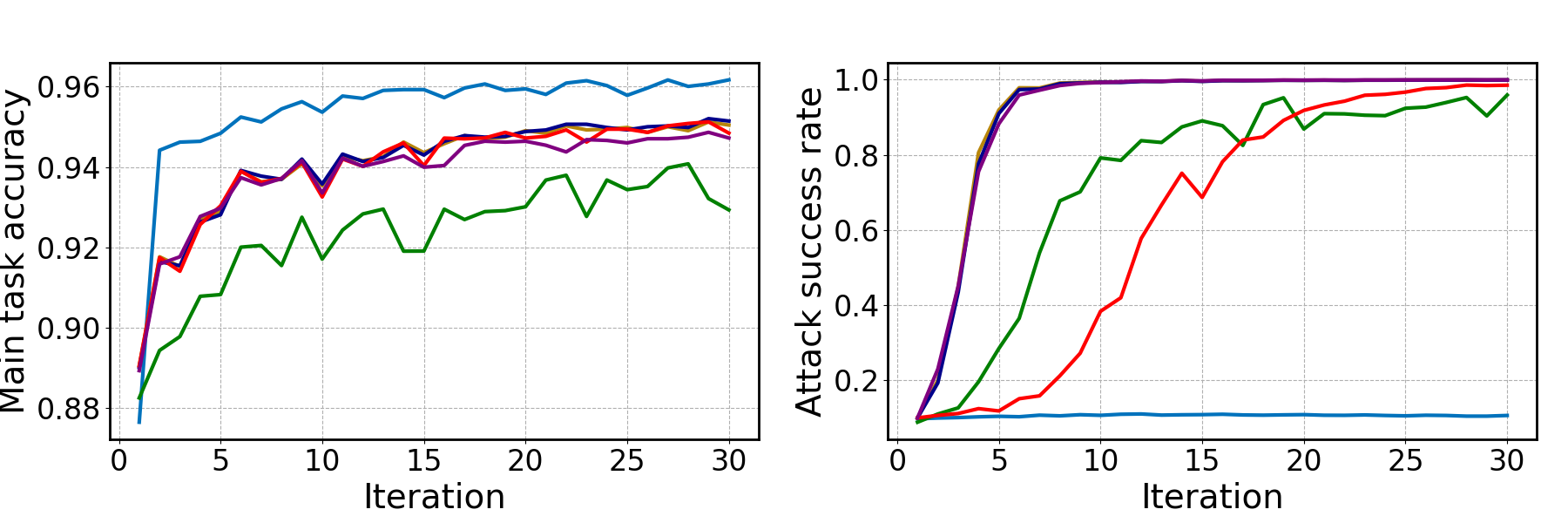 }
		\setlength{\belowcaptionskip}{0.1cm} \caption{LIE-5}
		\label{lie5}
	\end{subfigure}
		\centering
	\begin{subfigure}{0.49\linewidth}
		\centering
		\includegraphics[width=1\linewidth]{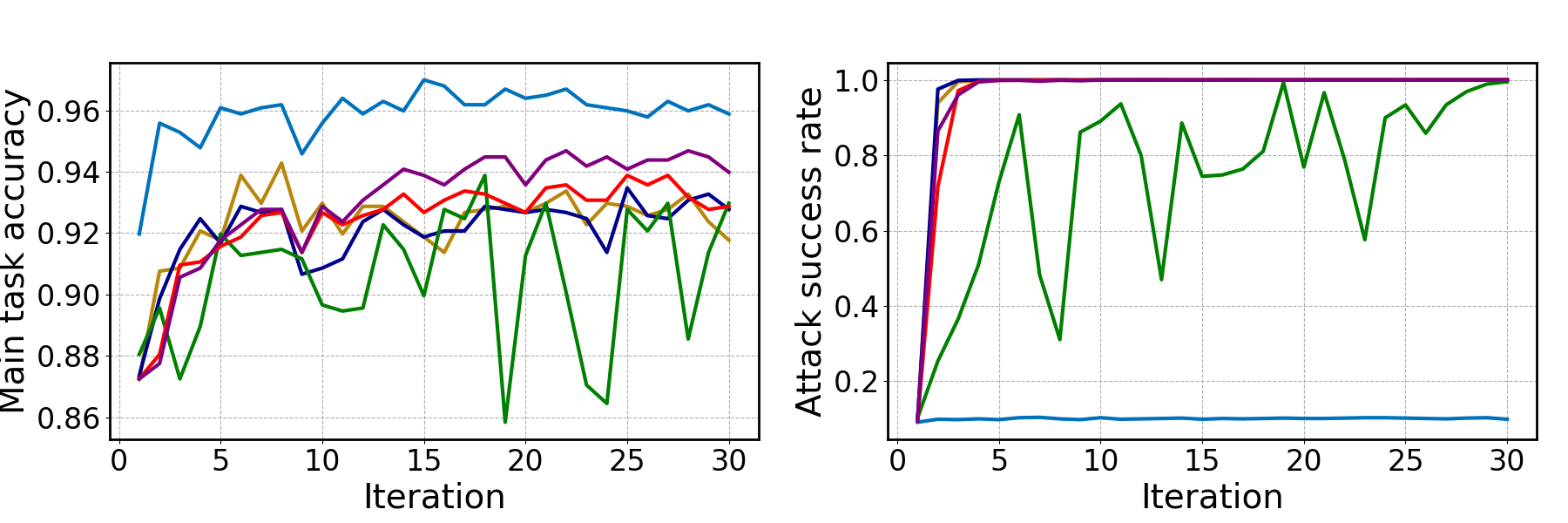}
		\setlength{\belowcaptionskip}{0.1cm} \caption{LIE-9}
		\label{lie9}
	\end{subfigure}
		\centering
	\begin{subfigure}{0.49\linewidth}
		\centering
		\includegraphics[width=1\linewidth]{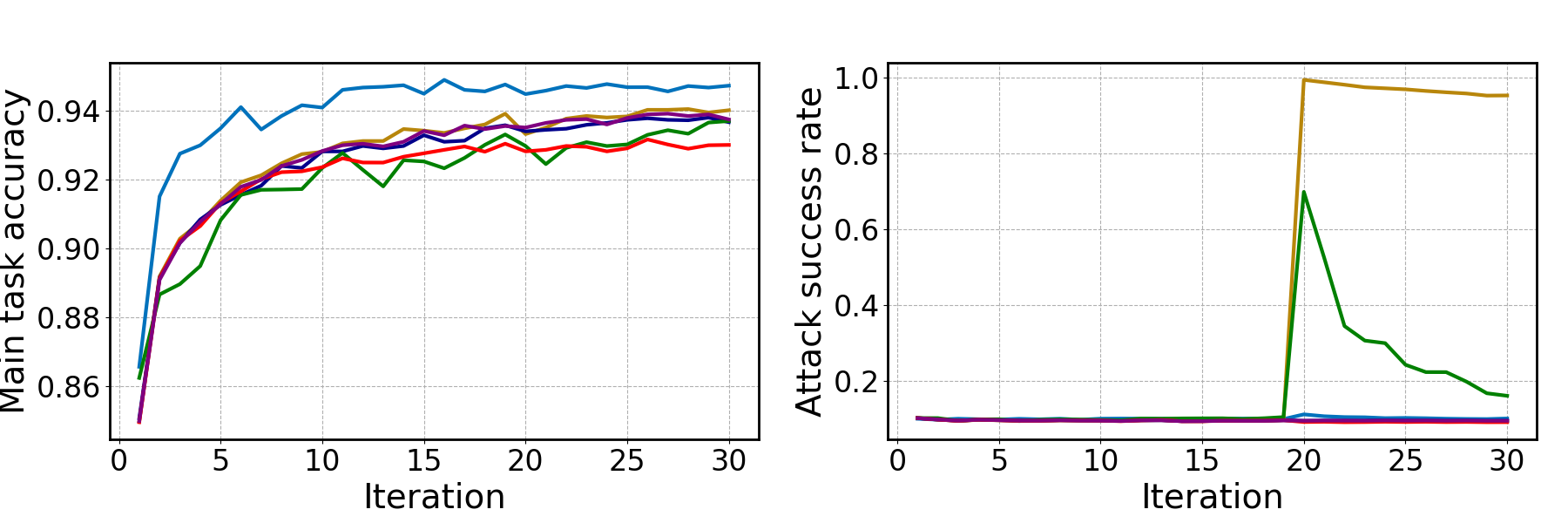}
		\setlength{\belowcaptionskip}{0.1cm} \caption{MPA-SS}
		\label{mpass}
	\end{subfigure}
		\centering
	\begin{subfigure}{0.49\linewidth}
		\centering
		\includegraphics[width=1\linewidth]{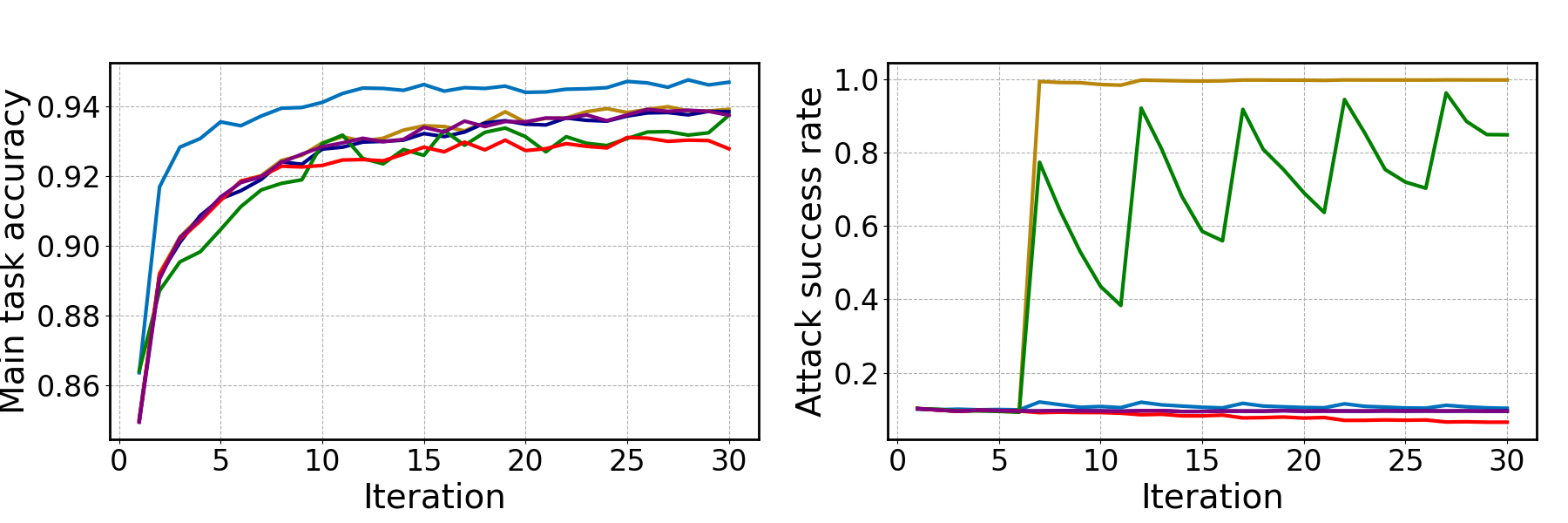}
		\setlength{\belowcaptionskip}{0.1cm} \caption{MPA-MS}
		\label{mpams}
	\end{subfigure}
 \centering
	\begin{subfigure}{0.49\linewidth}
		\centering
		\includegraphics[width=1\linewidth]{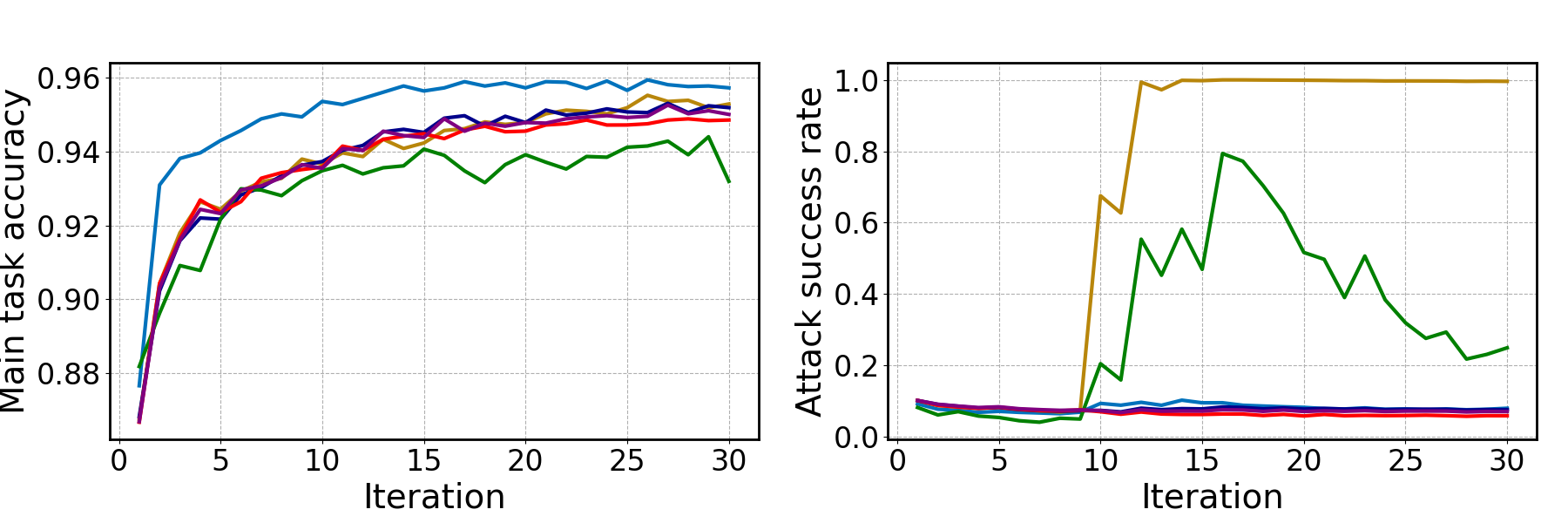}
		\setlength{\belowcaptionskip}{0.1cm} \caption{DBA-4*SS}
		\label{dba4ss}
	\end{subfigure}
 \centering
	\begin{subfigure}{0.49\linewidth}
		\centering
		\includegraphics[width=1\linewidth]{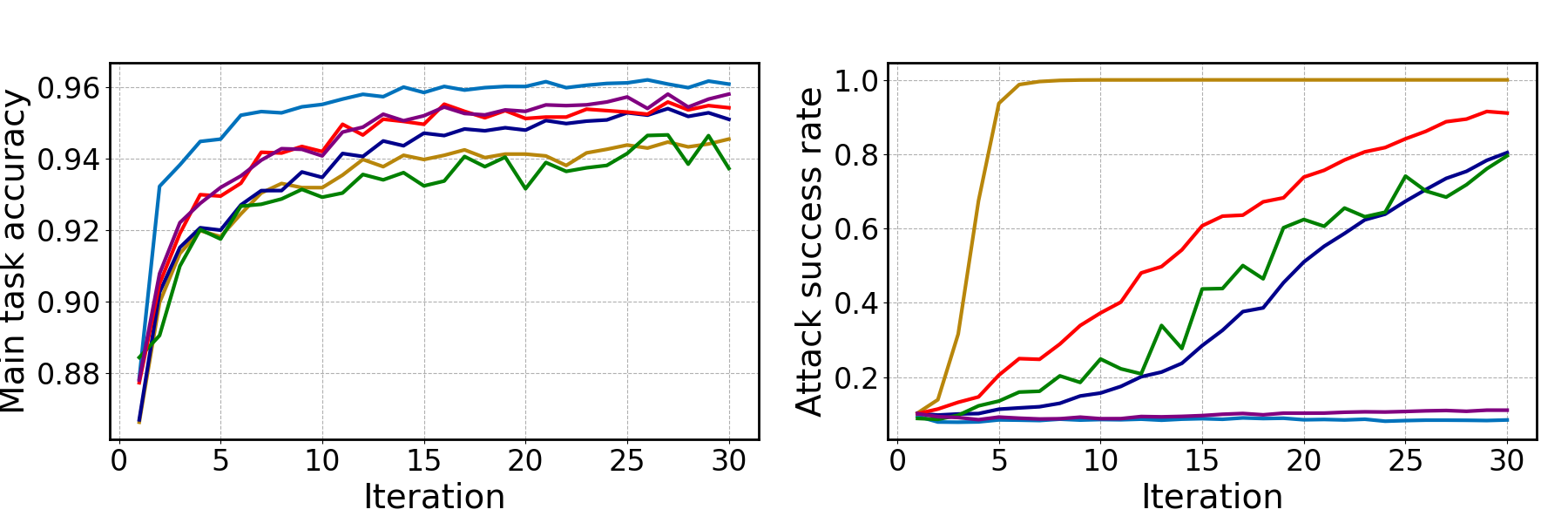}
		\setlength{\belowcaptionskip}{0.1cm} \caption{DBA-4*MS}
		\label{dba4ms}
	\end{subfigure}
 \centering
	\begin{subfigure}{0.49\linewidth}
		\centering
		\includegraphics[width=1\linewidth]{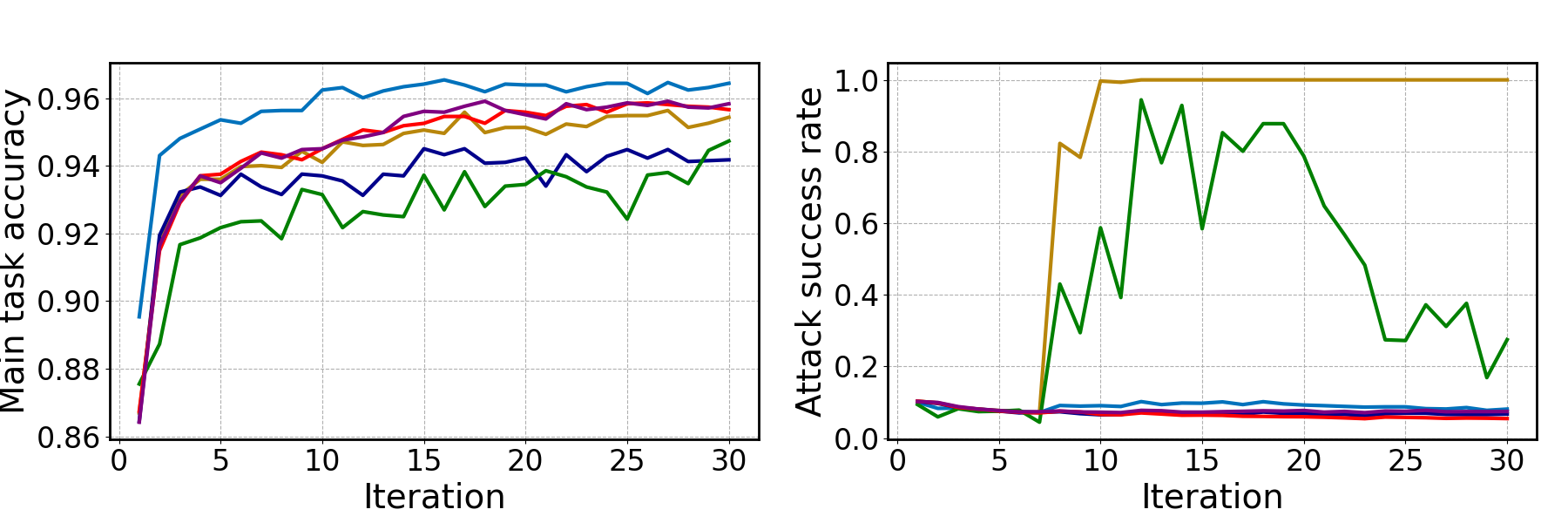}
		\setlength{\belowcaptionskip}{0.1cm} \caption{DBA-6*SS}
		\label{dba6ss}
	\end{subfigure}
 \centering
	\begin{subfigure}{0.49\linewidth}
		\centering
		\includegraphics[width=1\linewidth]{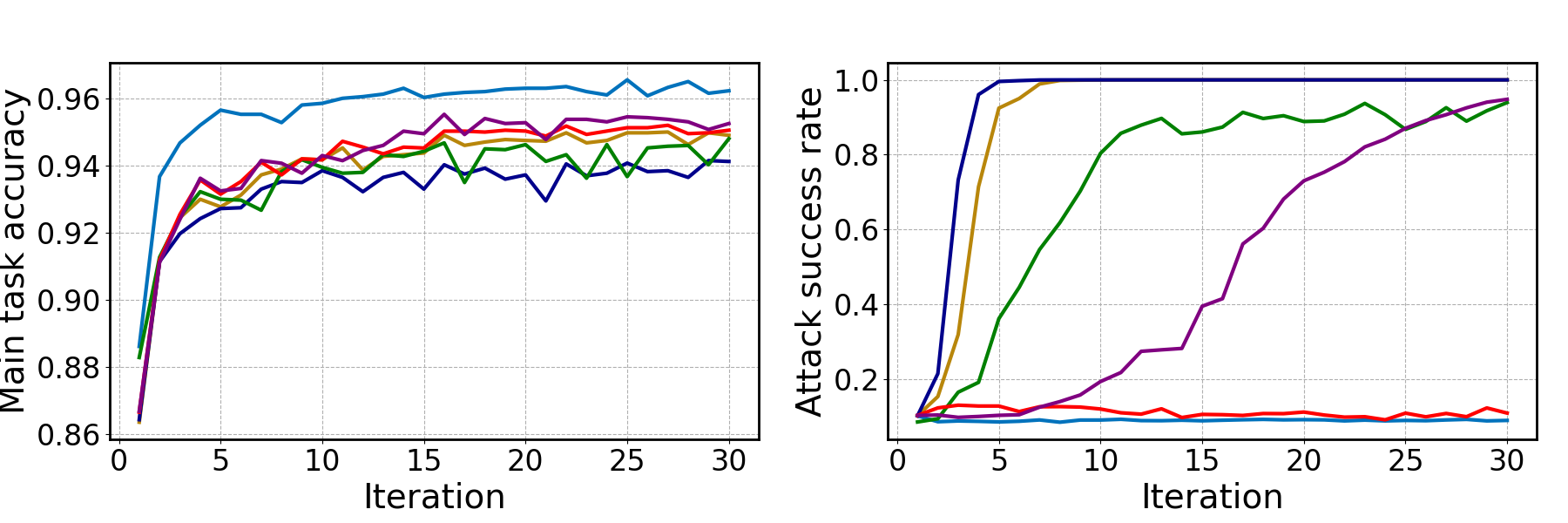}
		\setlength{\belowcaptionskip}{0.1cm} \caption{DBA-6*MS}
		\label{dba6ms}
	\end{subfigure}
	\captionsetup{font=small}\caption{Main task accuracy (left) and attack success rate (right) for canonical attacks on MNIST. Different defense methods are represented by different line colors given by the legend on the top-left subfigure. Each subfigure presents the results of one canonical attack: (a) DPA-5 and (b) DPA-9  with attack mounted at every iteration without scaling up the shared malicious model parameters; (c) LIE-5 and (d) LIE-9 with attack mounted at every iteration without scaling up the shared malicious model parameters; (e) MPA-SS with single-shot attack mounted at the 19-th iteration with a scaling factor of 10; (f) MPA-MS with attack mounted every 5 iterations with a scaling factor of 5; (g) DBA-4 and (i) DBA-6*SS with one-shot  attack mounted at the 9-th (7-th for DBA-6*SS), 11-th, 13-th and 15-th (17-th for DBA-6*SS) iterations with a scaling factor of 10; (h) DBA-4*MS and (j) DBA-6*MS with the attack mounted every iteration without scaling up the shared malicious model parameters. }
	\label{DE_MNIST}
 \end{figure*}
	
(1) SPFL is resilient against both model and data poisoning attacks, converges fast to very high MA while having very low ASR consistently against all attack methods. Other defenses can only reduce the ASR for certain attacks and attack settings, but were broken by at least one type of attack at some iteration as $N_m$ and $F_m$ increase. RFA and Median perform well against model poisoning attacks (MPA-SS, MPA-MS, DBA-4*SS and DBA-6*SS) where the magnitude of the malicious model parameters is significantly different from the benign ones, but fail to defend against colluding data poisoning attacks and other attacks without scale-up where the attack is performed every iteration by more than four adversaries (DPA-5, DPA-9, LIE-5, LIE-9 and DBA-6*MS). For all canonical attacks, SPFL maintains an ASR of around 0.1, which is as low as the fraction of test samples of the target class. For scaled-up attacks, its ASR is increased by at most 0.03 after the attack iteration, and reduces in subsequent training iterations.

(2) SPFL defends well against persistent attacks. For repeatedly mounted attacks, SPFL can still maintain a relatively constant ASR of around 0.1 with increasing iteration. Although FL-WBC can reduce the number of iterations over which the backdoor survives under single-shot MPA and DBA, its ASR increases gradually if these attacks are reinforced in subsequent training iterations. It fails to fully mitigate the backdoor effect against multiple-shot attacks.  

(3) SPFL can defend poisoning attack with almost no constraint on the number of colluding adversaries. The ASR of SPFL stays at around 0.1 even with $N_m = 9$. For both MNIST and CIFAR10, the ASR of RLR remains low at around 0.1 on only MPA and single-shot DBA attacks where only one adversary exists in a single iteration, as shown in subfigures (e), (f), (g) and (i). However, its ASR becomes as bad as FedAvg with multiple colluding adversaries shown in other subfigures except for DBA-6*MS attack in (j). Although six adversaries simultaneously exist in one attack round of BA-6*MS, the distributed triggers reduce the bias of the malicious model update or cause different model data distribution of each adversary, which keeps the aggregation of malicious model data within RLR’s defense bound. 
\begin{figure}[htbp]
	\centering
	\includegraphics[width=0.8\linewidth]{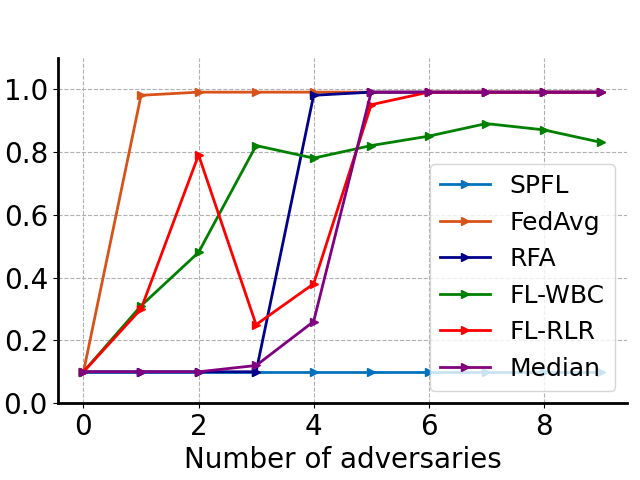}
	\captionsetup{font=small}\caption{Attack success rate for different number of adversaries on MNIST.}
	\label{number}
\end{figure}

We also evaluate the Byzantine tolerance of all defenses by increasing the number of sybils executing the DPA attack from 0 to 9 in Fig~\ref{number} for MNIST. One adversary would suffice to fully manipulate the model trained with FedAvg. The ASR of RFA increases drastically as the proportion of adversaries increases above 30$\%$. The ASR of Median increases slowly with $N_m$ until half of the clients are corrupted, at which point the defense breaks down abruptly. Similarly, RLR and FL-WBC can also withstand a limited fraction of sybils, but they cannot defend against a broad range of black-box attack configurations with a static set of parameters. The superior Byzantine tolerance of SPFL over them is attributed to the adaptive purification of adversarial perturbations. As the number of adversaries increases, the poisoning effect is stronger upon aggregation. The widening disparities between the student and teacher models in turn intensify the distillation of SPFL to forestall the extrapolation that jeopardizes the model integrity.  

 (4) SPFL can mitigate the influence of backdoor attacks while improving the main task accuracy. From Fig. \ref{DE_MNIST} and Fig. \ref{DE_CIFAR}, its MA is actually higher than that of FedAvg for each setting. Especially for the non-scaled up poisoning attack, the MA of its final trained model on MNIST is 0.96, which is up to 0.04 higher than that of FedAvg. On CIFAR10, SPFL exceeds the MA of FedAvg by a range from 0.01 to 0.03 over all canonical attacks. Other defense methods have varying degrees of sacrifice on the main task accuracy resulted from noises added onto the local model or biased aggregation rules. FL-WBC successfully mitigates  or eliminates  the influence of poisoning attack, but its MA has dropped  by up to 0.1 on CIFAR10. RLR also compromises accuracy, especially for defending against attacks with multiple adversaries. We also manually adjust $\theta$ to strengthen RLR for those attacks that its ASR is still high. As indicated by the red dashed line in Fig~\ref{DE_CIFAR}, the ASR is reduced with more significant sacrifice of MA. 

SPFL avoids sacrificing model quality by using KD from benign historical information to obliterate the malicious backdoor trigger from the aggregated model. At the same time, the student model also learns from its past cumulatively purified knowledge complementary to the knowledge learnt from other client models through aggregation. 
\begin{figure}[htbp]
	\centering
	\begin{subfigure}{0.49\linewidth}
		\centering
		\includegraphics[width=1\linewidth]{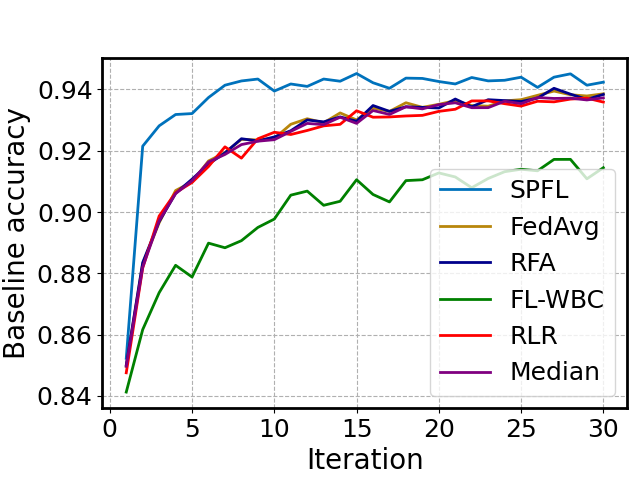}
		\caption{MNIST}
		\label{chutian3}
	\end{subfigure}
	\centering
	\begin{subfigure}{0.49\linewidth}
		\centering
		\includegraphics[width=1\linewidth]{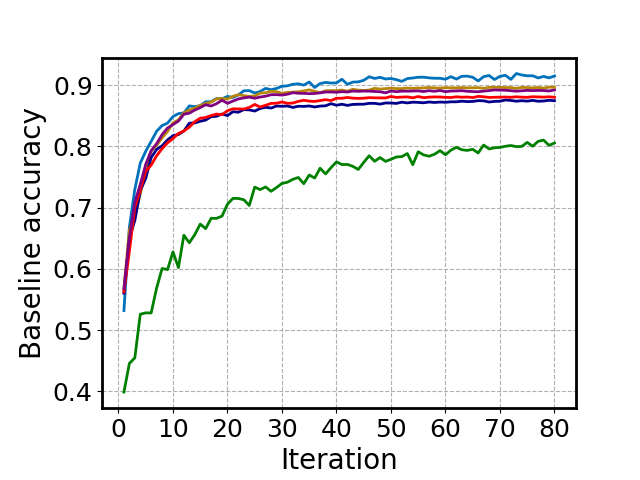}
		\caption{CIFAR10}
		\label{chutian3}
	\end{subfigure}
	\captionsetup{font=small}\caption{Benign accuracy comparison.}
	\label{BA}
\end{figure}

The BAs of all methods are compared in Fig. \ref{BA}. SPFL achieves the highest BA on both MNIST and CIFAR 10. It even exceeds the BA of FedAVG by 0.006 for MINST and 0.02 for CIFAR10 without increasing the number of communication rounds. SPFL has faster convergence and peak BA over FedAvg. Under the benign setting, Median can also achieve the same BA as FedAvg. The BAs of RFA and RLR trained models are well preserved for MNIST, but are dropped by approximately 0.02 for CIFAR10. The BA of FL-WBC is reduced by 0.02 for MNIST and more substantially by up to 0.1 for CIFAR10.

\subsection{Attention Attribution to SPFL’s Robustness}
SPFL transfers knowledge by gradient-weighted attention map, which highlights the important regions in the network's topology for an input prediction. To demonstrate the necessity of attention-based knowledge and the effectiveness of attention map construction in backdoor mitigation, we substitute the distillation loss function of SPFL with two contrasting loss functions for comparison. 

The first option, SPFL$\_$oA, omits the attention loss by transferring only logits-based knowledge without the attention-based loss for the student model, where the training loss is expressed as: \begin{equation}\label{eqa:ana}
    L_S = L_{CE,S} + \beta_{K,D}^t L_{KD,S}
\end{equation} 

The second option, SPFL$\_$NAD, calculates the attention map by a simple sum instead of a weighted combination of feature maps over all channels. Given the activation output of the $l$-th layer, $A^l \in R^{K\times U \times V}$, the attention map is represented as:

\begin{equation}\label{eqa:10}
    L(A^l) = \sum_{k=1}^K |A_k|
\end{equation}  
\vspace{-0.6cm}
\begin{figure}[htbp]
	\centering
	\begin{subfigure}{0.98\linewidth}
    \setlength\abovecaptionskip{-0.2\baselineskip}
		\centering
		\includegraphics[width=1\linewidth]{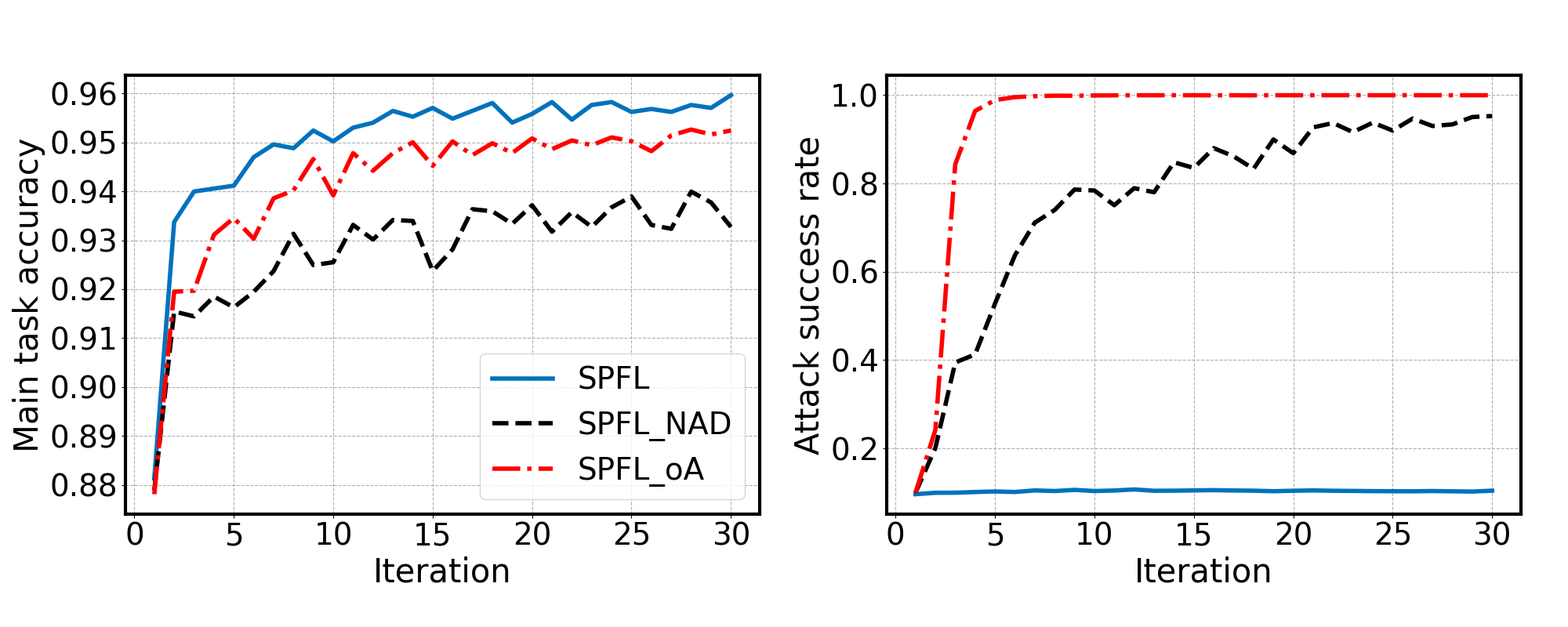}
		\caption{MNIST}
		\label{chutian3}
	\end{subfigure}
	\centering
	\begin{subfigure}{0.98\linewidth}
    \setlength\abovecaptionskip{-0.2\baselineskip}
		\centering
		\includegraphics[width=1\linewidth]{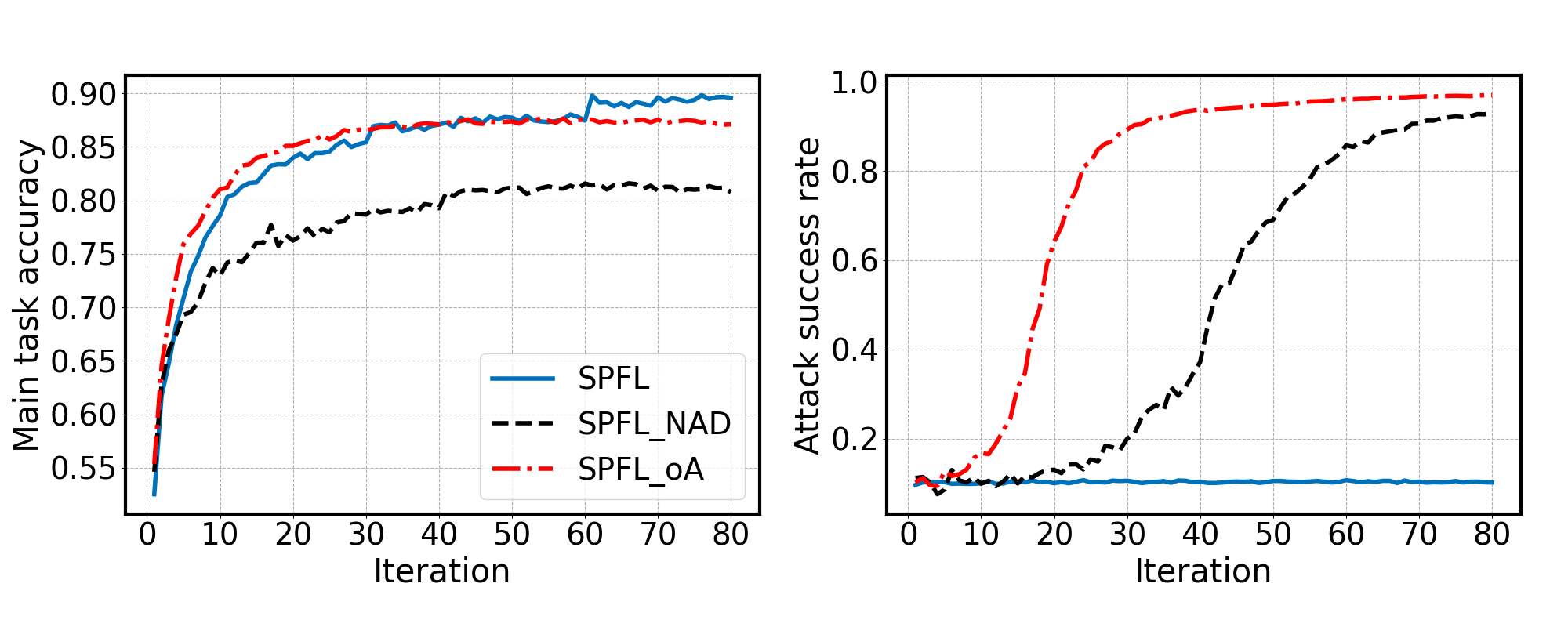}
		\caption{CIFAR10}
		\label{chutian3}
	\end{subfigure}
	\captionsetup{font=small}\caption{MAs and ASRs of SPFL with different distillation loss functions against DPA-5 attack. }
	\label{loss function}
\end{figure}
\begin{figure}[!ht]	
\centering
	\begin{subfigure}{0.24\linewidth}
		\centering
		\includegraphics[width=1\linewidth]{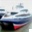}
	\end{subfigure}
	\centering
	\begin{subfigure}{0.24\linewidth}
		\centering
		\includegraphics[width=1\linewidth]{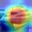}
	\end{subfigure}
	\centering
	\begin{subfigure}{0.24\linewidth}
		\centering
		\includegraphics[width=1\linewidth]{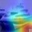}
		
	\end{subfigure}
		\centering
	\begin{subfigure}{0.24\linewidth}
		\centering
		\includegraphics[width=1\linewidth]{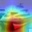}
			\end{subfigure}
	
		\centering
	\begin{subfigure}{0.24\linewidth}
		\centering
		\includegraphics[width=1\linewidth]{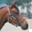}
		
	\end{subfigure}
	\centering
	\begin{subfigure}{0.24\linewidth}
		\centering
		\includegraphics[width=1\linewidth]{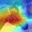}
			\end{subfigure}
	\centering
	\begin{subfigure}{0.24\linewidth}
		\centering
		\includegraphics[width=1\linewidth]{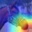}
		
	\end{subfigure}
		\centering
	\begin{subfigure}{0.24\linewidth}
		\centering
		\includegraphics[width=1\linewidth]{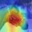}
		
	\end{subfigure}
	
		\centering
	\begin{subfigure}{0.24\linewidth}
		\centering
		\includegraphics[width=1\linewidth]{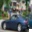}
		\caption{}
		
	\end{subfigure}
	\centering
	\begin{subfigure}{0.24\linewidth}
		\centering
		\includegraphics[width=1\linewidth]{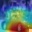}
		\caption{}
		
	\end{subfigure}
	\centering
	\begin{subfigure}{0.24\linewidth}
		\centering
		\includegraphics[width=1\linewidth]{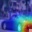}
		\caption{}
			\end{subfigure}
		\centering
	\begin{subfigure}{0.24\linewidth}
		\centering
		\includegraphics[width=1\linewidth]{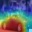}
		\caption{}
			\end{subfigure}
	
	\captionsetup{font=small}\caption{Attention maps (AM) of SPFL trained model: (a) Input images with backdoor trigger; (b) AM of  teacher model; (c) AM of student model before distillation; (d) AM of student model after distillation.}
	\label{AM_SPFL}
\end{figure}

The MA and ASR against DPA-5 attack of the student model trained with these alternative loss functions are compared with those of SPFL in Fig~\ref{loss function}. SPFL outperforms these alternative distillation options in suppressing the backdoor and preserving the main task accuracy. The ASR of the student model trained with SPFL$\_$oA increases with iterations to the peak value (i.e., 1 for MNIST and 0.97 for CIFAR10) but its MA is lower than SPFL by 0.01. This may be attributed to the reduced information possessed by logits. Therefore, the logits of clean inputs are less affected by the backdoor bias. SPFL$\_$NAD weakens the influence of backdoor to a certain extent. However, as the attack iterates, its ASR is higher than 0.9 eventually for both MNIST and CIFAR10. The MA of the trained model has been dropped substantially by a difference of up to 0.1 for CIFAR10. Unlike SPFL$\_$NAD, which combines feature maps over all channels evenly, SPFL performs a weighted combination of forward feature maps, where the weight is derived from the gradient of the score for ground-truth class with respect to the corresponding feature maps. This way, more knowledge is distilled from feature maps that contribute most to the correct prediction.

\begin{figure*}[htbp]
	\centering
	\begin{subfigure}{0.13\textwidth}
  	\centering
		\includegraphics[width=1\linewidth]{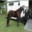}
		\label{chutian3}
	\end{subfigure}
	\centering
	\begin{subfigure}{0.13\textwidth}
		\centering
		\includegraphics[width=1\linewidth]{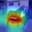}
		\label{chutian3}
	\end{subfigure}
	\centering
	\begin{subfigure}{0.13\textwidth}
		\centering
		\includegraphics[width=1\linewidth]{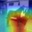}
		\label{chutian3}
	\end{subfigure}
		\centering
	\begin{subfigure}{0.13\textwidth}
		\centering
		\includegraphics[width=1\linewidth]{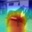}
		\label{AM}
	\end{subfigure}
	\centering
	\begin{subfigure}{0.13\textwidth}
		\centering
		\includegraphics[width=1\linewidth]{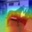}
		\label{chutian3}
	\end{subfigure}
	\centering
	\begin{subfigure}{0.13\textwidth}
		\centering
		\includegraphics[width=1\linewidth]{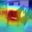}
		\label{chutian3}
	\end{subfigure}
	\centering
	\begin{subfigure}{0.13\textwidth}
		\centering
		\includegraphics[width=1\linewidth]{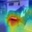}
		\label{chutian3}
	\end{subfigure}
	\centering
	\begin{subfigure}{0.13\textwidth}
		\centering
		\includegraphics[width=1\linewidth]{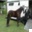}
		\caption{Original image}
		\label{AM}
	\end{subfigure}
    \centering
	\begin{subfigure}{0.13\textwidth}
		\centering
		\includegraphics[width=1\linewidth]{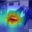}
		\caption{SPFL}
		\label{chutian3}
	\end{subfigure}
	\centering
	\begin{subfigure}{0.13\textwidth}
		\centering
		\includegraphics[width=1\linewidth]{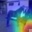}
 		\caption{FedAvg}
		\label{chutian3}
	\end{subfigure}
	\centering
	\begin{subfigure}{0.13\textwidth}
		\centering
		\includegraphics[width=1\linewidth]{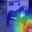}
		\caption{RFA}
		\label{AM}
	\end{subfigure}
	\centering
	\begin{subfigure}{0.13\textwidth}
		\centering
		\includegraphics[width=1\linewidth]{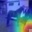}
		\caption{Median }
		\label{chutian3}
	\end{subfigure}
	\centering
	\begin{subfigure}{0.13\textwidth}
		\centering
		\includegraphics[width=1\linewidth]{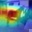}
		\caption{RLR}
		\label{chutian3}
	\end{subfigure}
	\centering
	\begin{subfigure}{0.13\textwidth}
		\centering
		\includegraphics[width=1\linewidth]{0-20CAM_WBC.jpg}
		\caption{FL-WBC}
		\label{chutian3}
	\end{subfigure}
	\captionsetup{font=small}\caption{Visualization of attention maps (AM) of models trained with different methods: (a) Original image; (b)-(g) AM of models trained with SPFL, FedAvg, RFA, Median, RLR and FL-WBC. Results for clean and poisoned images are displayed on the top and bottom rows, respectively.}
	\label{AM_S}
\vspace{-0.4cm}
\end{figure*}

Fig.~\ref{AM_SPFL} compares the attention maps of the teacher model (b) and the student model before and after distillation ((c) and (d)) for three poisoned images with a white square trigger at the lower right corner (a). As depicted in Fig.~\ref{AM_SPFL}(c), the attention maps of the backdoored student model before distillation are completely biased towards the trigger region, causing misprediction. After distillation training supervised by the local teacher model (Fig.~\ref{AM_SPFL}(b)), the attention has been corrected to the region for correct prediction in Fig.~\ref{AM_SPFL}(d).

Fig.~\ref{AM_S} compares the attention maps obtained from the trained models of SPFL, FedAvg, RFA, Median, RLR and FL-WBC under LIE-5  attack. For the models trained with FedAvg, RFA and Median, the attention maps on the clean image (top row) differ significantly from their attention maps of the poisoned image (bottom row). The attention of these trained models has been biased to the region where the trigger pattern resides. On the contrary, the attention maps on input with trigger are not shifted apparently for the federated models trained by SPFL, RLR and Fed-WBC. This explains why the LIE-5 attack was not successful on these methods in the earlier experiments. The average distance between attention maps of a trained federated model on clean input samples and on the corresponding backdoored samples can be used to quantify the effectiveness of mitigation. The smaller the distance between them, the less likely the backdoor can be successfully activated. This average distance for all samples in the test dataset is calculated for each defense method against LIE. The distances for SPFL, FedAvg, RFA, Median, RLR and FL-WBC are 2.09, 9.91, 10.03, 11.12 6.06 and 3.41, respectively. The smallest value of SPFL corroborates its greatest resistance to poisoning attacks.

\section{Conclusion}
This paper presents a robust client-based defense against poising attacks in FL setting. It leverages attention-based self-distillation to independently purify the aggregated model received by each benign client to deliver credible performance guarantee. Our experimental results show that SPFL can mitigate a variety of poisoning attacks even if the number of adversaries and the attack frequency exceed the threshold defendable by state-of-the-art methods. Unlike existing defenses, SPFL raises instead of trades main task accuracy for security. As the self-distillation training needs only to be performed by the benign clients locally, SPFL is drop-in compatible and complementary with any aggregation methods, communication protocols and privacy-preserving protection schemes applicable to standard FL settings. It can be easily deployed to increase the Byzantine tolerance of a FL system without sacrificing the original quality and advantages attainable by other coexisting components. 


%

\ifCLASSOPTIONcaptionsoff
  \newpage
\fi



%
\bibliographystyle{IEEEtran}
\bibliography{IEEEabrv}

\begin{thebibliography}{10}
\providecommand{\url}[1]{#1}
\csname url@samestyle\endcsname
\providecommand{\newblock}{\relax}
\providecommand{\bibinfo}[2]{#2}
\providecommand{\BIBentrySTDinterwordspacing}{\spaceskip=0pt\relax}
\providecommand{\BIBentryALTinterwordstretchfactor}{4}
\providecommand{\BIBentryALTinterwordspacing}{\spaceskip=\fontdimen2\font plus
\BIBentryALTinterwordstretchfactor\fontdimen3\font minus
  \fontdimen4\font\relax}
\providecommand{\BIBforeignlanguage}[2]{{%
\expandafter\ifx\csname l@#1\endcsname\relax
\typeout{** WARNING: IEEEtran.bst: No hyphenation pattern has been}%
\typeout{** loaded for the language `#1'. Using the pattern for}%
\typeout{** the default language instead.}%
\else
\language=\csname l@#1\endcsname
\fi
#2}}
\providecommand{\BIBdecl}{\relax}
\BIBdecl

\bibitem{yang2019federated}
Q.~Yang, Y.~Liu, T.~Chen, and Y.~Tong, ``Federated machine learning: Concept
  and applications,'' \emph{ACM Trans. on Intell. Syst. and Technol.}, vol.~10,
  no.~2, pp. 1--19, 2019.

\bibitem{41zhu2019deep}
L.~Zhu, Z.~Liu, and S.~Han, ``Deep leakage from gradients,'' \emph{Advances in
  Neural Inform. Process. Syst.}, vol.~32, 2019.

\bibitem{20dp}
K.~Wei \emph{et~al.}, ``Federated learning with differential privacy:
  Algorithms and performance analysis,'' \emph{IEEE Trans. on Inform. Forensics
  and Secur.}, vol.~15, pp. 3454--3469, Apr. 2020.

\bibitem{1}
E.~Bagdasaryan, A.~Veit, Y.~Hua, D.~Estrin, and V.~Shmatikov, ``How to backdoor
  federated learning,'' in \emph{Proc. Int. Conf. on Artif. Intell. and
  Statist.}\hskip 1em plus 0.5em minus 0.4em\relax PMLR, 2020, pp. 2938--2948.

\bibitem{blanchard2017machine}
P.~Blanchard, E.~M. El~Mhamdi, R.~Guerraoui, and J.~Stainer, ``Machine learning
  with adversaries: Byzantine tolerant gradient descent,'' \emph{Advances in
  Neural Inform. Process. Syst.}, vol.~30, 2017.

\bibitem{yin2018byzantine}
D.~Yin, Y.~Chen, R.~Kannan, and P.~Bartlett, ``Byzantine-robust distributed
  learning: Towards optimal statistical rates,'' in \emph{Proc. Int. Conf. on
  Mach. Learn.}\hskip 1em plus 0.5em minus 0.4em\relax PMLR, 2018, pp.
  5650--5659.

\bibitem{pillutla2022robust}
K.~Pillutla, S.~M. Kakade, and Z.~Harchaoui, ``Robust aggregation for federated
  learning,'' \emph{IEEE Trans. on Signal Process.}, vol.~70, pp. 1142--1154,
  2022.

\bibitem{18fung2018mitigating}
C.~Fung, C.~J. Yoon, and I.~Beschastnikh, ``The limitations of federated
  learning in sybil settings,'' in \emph{Proc. Int. Symp. on Res. in Attacks,
  Intrusions and Defenses}, 2020, pp. 301--316.

\bibitem{chang2019cronus}
H.~Chang, V.~Shejwalkar, R.~Shokri, and A.~Houmansadr, ``Cronus: Robust and
  heterogeneous collaborative learning with black-box knowledge transfer,''
  \emph{arXiv preprint arXiv:1912.11279}, 2019.

\bibitem{40kairouz2021advances}
P.~Kairouz \emph{et~al.}, ``Advances and open problems in federated learning,''
  \emph{Found. and Trends{\textregistered} in Mach. Learn.}, vol.~14, no. 1--2,
  pp. 1--210, 2021.

\bibitem{43bonawitz2017practical}
K.~Bonawitz \emph{et~al.}, ``Practical secure aggregation for
  privacy-preserving machine learning,'' in \emph{Proc. ACM SIGSAC Conf. on
  Comput. and Commun. Secur.}, 2017, pp. 1175--1191.

\bibitem{aono2017privacy}
Y.~Aono \emph{et~al.}, ``Privacy-preserving deep learning via additively
  homomorphic encryption,'' \emph{IEEE Trans. on Inf. Forensics and Secur.},
  vol.~13, no.~5, pp. 1333--1345, 2017.

\bibitem{liu2022dhsa}
Z.~Liu \emph{et~al.}, ``{DHSA}: efficient doubly homomorphic secure aggregation
  for cross-silo federated learning,'' \emph{J Supercomput.}, pp. 1--31, 2022.

\bibitem{liu2022sash}
{Z. Liu} \emph{et~al.}, ``{SASH}: Efficient secure aggregation based on shprg
  for federated learning,'' in \emph{Proc. Conf. Uncertainty Artif.
  Intell.}\hskip 1em plus 0.5em minus 0.4em\relax PMLR, 2022, pp. 1243--1252.

\bibitem{shejwalkar2021manipulating}
V.~Shejwalkar and A.~Houmansadr, ``Manipulating the byzantine: Optimizing model
  poisoning attacks and defenses for federated learning,'' in \emph{Proc. Netw.
  and Distrib. Syst. Secur. Symp.}, 2021.

\bibitem{5}
C.~Xie, K.~Huang, P.-Y. Chen, and B.~Li, ``{DBA}: Distributed backdoor attacks
  against federated learning,'' in \emph{Proc. Int. Conf. on Learn.
  Representations}, Apr. 2020.

\bibitem{fang2020local}
M.~Fang, X.~Cao, J.~Jia, and N.~Gong, ``Local model poisoning attacks to
  byzantine-robust federated learning,'' in \emph{Proc. 29th USENIX Secur.
  Symp.}, Aug. 2020, pp. 1605--1622.

\bibitem{2}
Z.~Sun, P.~Kairouz, A.~T. Suresh, and H.~B. McMahan, ``Can you really backdoor
  federated learning?'' \emph{arXiv preprint arXiv:1911.07963}, 2019.

\bibitem{6baruch2019little}
G.~Baruch, M.~Baruch, and Y.~Goldberg, ``A little is enough: Circumventing
  defenses for distributed learning,'' \emph{Advances in Neural Inform.
  Process. Syst.}, vol.~32, pp. 8645--8645, 2019.

\bibitem{3}
A.~N. Bhagoji, S.~Chakraborty, P.~Mittal, and S.~Calo, ``Analyzing federated
  learning through an adversarial lens,'' in \emph{Proc. Int. Conf. on Mach.
  Learn.}\hskip 1em plus 0.5em minus 0.4em\relax PMLR, 2019, pp. 634--643.

\bibitem{37cao2020fltrust}
X.~Cao, M.~Fang, J.~Liu, and N.~Z. Gong, ``{FLTrust}: Byzantine-robust
  federated learning via trust bootstrapping,'' in \emph{Proc. Netw. and
  Distrib. Syst. Secur. Symp.}, 2021.

\bibitem{8munoz2019byzantine}
L.~Mu{\~n}oz-Gonz{\'a}lez, K.~T. Co, and E.~C. Lupu, ``Byzantine-robust
  federated machine learning through adaptive model averaging,'' \emph{arXiv
  preprint arXiv:1909.05125}, 2019.

\bibitem{9ozdayi2021defending}
M.~S. Ozdayi, M.~Kantarcioglu, and Y.~R. Gel, ``Defending against backdoors in
  federated learning with robust learning rate,'' in \emph{Proc. of the AAAI
  Conf. on Artif. Intell.}, vol.~35, no.~10, 2021, pp. 9268--9276.

\bibitem{shieldFL22}
Z.~Ma, J.~Ma, Y.~Miao, Y.~Li, and R.~H. Deng, ``{ShieldFL}: Mitigating model
  poisoning attacks in privacy-preserving federated learning,'' \emph{IEEE
  Trans. on Inform. Forensics and Secur.}, vol.~17, pp. 1639--1654, Apr. 2022.

\bibitem{7li2020learning}
S.~Li, Y.~Cheng, W.~Wang, Y.~Liu, and T.~Chen, ``Learning to detect malicious
  clients for robust federated learning,'' \emph{arXiv preprint
  arXiv:2002.00211}, 2020.

\bibitem{rieger2022deepsight}
P.~Rieger, T.~D. Nguyen, M.~Miettinen, and A.-R. Sadeghi, ``Deepsight:
  Mitigating backdoor attacks in federated learning through deep model
  inspection,'' \emph{arXiv preprint arXiv:2201.00763}, 2022.

\bibitem{11shen2016auror}
S.~Shen, S.~Tople, and P.~Saxena, ``Auror: Defending against poisoning attacks
  in collaborative deep learning systems,'' in \emph{Proc. 32nd Annual Conf. on
  Comput. Secur. Appl.}, 2016, pp. 508--519.

\bibitem{12andreina2021baffle}
S.~Andreina, G.~A. Marson, H.~M{\"o}llering, and G.~Karame, ``Baffle: Backdoor
  detection via feedback-based federated learning,'' in \emph{Proc. IEEE 41st
  Int. Conf. on Distrib. Comput. Syst.}, Jul. 2021, pp. 852--863.

\bibitem{19sun2021fl}
J.~Sun \emph{et~al.}, ``{FL-WBC}: Enhancing robustness against model poisoning
  attacks in federated learning from a client perspective,'' \emph{Advances in
  Neural Inform. Process. Syst.}, vol.~34, pp. 12\,613--12\,624, 2021.

\bibitem{17chen2020backdoor}
C.-L. Chen, L.~Golubchik, and M.~Paolieri, ``Backdoor attacks on federated
  meta-learning,'' \emph{arXiv preprint arXiv:2006.07026}, 2020.

\bibitem{20zhang2020defending}
J.~Zhang, D.~Wu, C.~Liu, and B.~Chen, ``Defending poisoning attacks in
  federated learning via adversarial training method,'' in \emph{Proc. Int.
  Conf. on Frontiers in Cyber Secur.}\hskip 1em plus 0.5em minus 0.4em\relax
  Springer, 2020, pp. 83--94.

\bibitem{tramer2016stealing}
F.~Tram{\`e}r, F.~Zhang, A.~Juels, M.~K. Reiter, and T.~Ristenpart, ``Stealing
  machine learning models via prediction apis,'' in \emph{Proc. 25th USENIX
  Secur. Symp.}, 2016, pp. 601--618.

\bibitem{22hinton2015distilling}
G.~Hinton, O.~Vinyals, J.~Dean \emph{et~al.}, ``Distilling the knowledge in a
  neural network,'' \emph{arXiv preprint arXiv:1503.02531}, vol.~2, no.~7,
  2015.

\bibitem{24zhang2019your}
L.~Zhang \emph{et~al.}, ``Be your own teacher: Improve the performance of
  convolutional neural networks via self distillation,'' in \emph{Proc.
  IEEE/CVF Int. Conf. Comput. Vision}, 2019, pp. 3713--3722.

\bibitem{27papernot2016distillation}
N.~Papernot, P.~McDaniel, X.~Wu, S.~Jha, and A.~Swami, ``Distillation as a
  defense to adversarial perturbations against deep neural networks,'' in
  \emph{Proc. IEEE Symp. on Secur. and Privacy}, 2016, pp. 582--597.

\bibitem{lee2019rethinking}
H.~Lee, S.~Hwang, and J.~Shin, ``Rethinking data augmentation: Self-supervision
  and self-distillation,'' \emph{arXiv preprint arXiv:1910.05872}, 2019.

\bibitem{34zhang2018deep}
Y.~Zhang, T.~Xiang, T.~M. Hospedales, and H.~Lu, ``Deep mutual learning,'' in
  \emph{Proc. IEEE conf. comput. vision and pattern recognit.}, 2018, pp.
  4320--4328.

\bibitem{35chen2020online}
D.~Chen, J.-P. Mei, C.~Wang, Y.~Feng, and C.~Chen, ``Online knowledge
  distillation with diverse peers,'' in \emph{Proc. AAAI Conf. on Artif.
  Intell.}, vol.~34, no.~4, 2020, pp. 3430--3437.

\bibitem{36anil2018large}
R.~Anil \emph{et~al.}, ``Large scale distributed neural network training
  through online distillation,'' in \emph{Proc. Int. Conf. on Learn.
  Representations}, 2018.

\bibitem{25hou2019learning}
Y.~Hou, Z.~Ma, C.~Liu, and C.~C. Loy, ``Learning lightweight lane detection
  cnns by self attention distillation,'' in \emph{Proc. IEEE/CVF Int. Conf.
  Comput. Vision}, 2019, pp. 1013--1021.

\bibitem{26yang2019snapshot}
C.~Yang, L.~Xie, C.~Su, and A.~L. Yuille, ``Snapshot distillation:
  Teacher-student optimization in one generation,'' in \emph{Proc. IEEE/CVF
  Conf. Comput. Vision and Pattern Recognition}, 2019, pp. 2859--2868.

\bibitem{li2021neural}
Y.~Li \emph{et~al.}, ``Neural attention distillation: Erasing backdoor triggers
  from deep neural networks,'' in \emph{Proc. Int. Conf. on Learn.
  Representations}, 2021.

\bibitem{45}
H.~Yu, S.~Yang, and S.~Zhu, ``Parallel restarted sgd with faster convergence
  and less communication: Demystifying why model averaging works for deep
  learning,'' in \emph{Proc. of the AAAI Conf. on Artificial Intelligence},
  vol.~33, Jul. 2019, pp. 5693--5700.

\bibitem{47back}
V.~Shejwalkar, A.~Houmansadr, P.~Kairouz, and D.~Ramage, ``Back to the drawing
  board: A critical evaluation of poisoning attacks on production federated
  learning,'' in \emph{Proc. IEEE Symp. on Secur. and Privacy}, 2022, pp.
  1354--1371.

\bibitem{48}
M.~Nwadike, T.~Miyawaki, E.~Sarkar, M.~Maniatakos, and F.~Shamout,
  ``Explainability matters: Backdoor attacks on medical imaging,'' \emph{arXiv
  preprint arXiv:2101.00008}, 2020.

\bibitem{mnist}
L.~Deng, ``The mnist database of handwritten digit images for machine learning
  research [best of the web],'' \emph{IEEE Signal Process. Mag.}, vol.~29,
  no.~6, pp. 141--142, 2012.

\bibitem{resnet}
K.~He, X.~Zhang, S.~Ren, and J.~Sun, ``Deep residual learning for image
  recognition,'' \emph{CoRR: abs/1512.03385}, 2015.

\bibitem{2012Learning}
A.~Krizhevsky, G.~Hinton \emph{et~al.}, ``Learning multiple layers of features
  from tiny images,'' \emph{University of Toronto, Canada}, 2009.

\bibitem{gu2017badnets}
T.~Gu, B.~Dolan-Gavitt, and S.~Garg, ``Badnets: Identifying vulnerabilities in
  the machine learning model supply chain,'' \emph{arXiv preprint
  arXiv:1708.06733}, 2017.

\bibitem{lyu2022privacy}
L.~Lyu \emph{et~al.}, ``Privacy and robustness in federated learning: Attacks
  and defenses,'' \emph{IEEE Trans. on Neural Netw. and Learn. Syst.}, vol.~99,
  pp. 1--21, Nov. 2022.

\end{thebibliography}

%

\end{document}